\documentclass[structabstract]{aa}  
\usepackage{txfonts,graphicx,natbib}
\begin{document}
\title{Properties of dust in the Galactic center region probed by AKARI far-infrared spectral mapping -- detection of a dust feature}
\titlerunning{Dust properties in the Galactic center}
\authorrunning{H. Kaneda et al.}
\author{H. Kaneda\inst{1}, A. Yasuda\inst{1,2}, T. Onaka\inst{3}, M. Kawada\inst{2}, N. Murakami\inst{6}, T. Nakagawa\inst{2}, Y. Okada\inst{4}, H. Takahashi\inst{5}}
\institute{
Graduate School of Science, Nagoya University, Chikusa-ku, Nagoya, 464-8602, Japan \\
\email{kaneda@u.phys.nagoya-u.ac.jp}\\
\and
Institute of Space and Astronautical Science, Japan Aerospace Exploration Agency, Sagamihara, Kanagawa 252-5210, Japan\\
\and
Department of Astronomy, Graduate School of Science, University of Tokyo, \\
Bunkyo-ku, Tokyo 113-0003, Japan\\
\and
I. Physikalisches Institut, Universit\"at zu K\"oln, Z\"ulpicher Str. 77, 50937 K\"oln, Germany\\
\and
Institute of Astronomy, Graduate School of Science, University of Tokyo, \\
Mitaka, Tokyo 181-0015, Japan\\
\and
Bisei Astronomical Observatory, Ibara, Okayama 714-1411, Japan
}

\date{Received; accepted}

\abstract
{}
{We investigate the properties of interstellar dust in the Galactic center region toward the Arches and Quintuplet clusters.}
{With the Fourier Transform Spectrometer of the AKARI/Far-Infrared Surveyor, we performed the far-infrared (60-140~cm$^{-1}$) spectral mapping of an area of about 10$'$~$\times$~10$'$ which includes the two clusters to obtain a low-resolution ($R=1.2$ cm$^{-1}$) spectrum at every spatial bin of $30''\times 30''$.}
{We derive the spatial variations of dust continuum emission at different wavenumbers, which are compared with those of the [O{\small III}] 88~$\mu$m (113~cm$^{-1}$) emission and the OH~119~$\mu$m (84~cm$^{-1}$) absorption. The spectral fitting shows that two dust modified blackbody components with temperatures of $\sim$20~K and $\sim$50~K can reproduce most of the continuum spectra. For some spectra, however, we find that there exists a significant excess on top of a modified blackbody continuum around 80--90~cm$^{-1}$ (110--130~$\mu$m).} 
{The warmer dust component is spatially correlated well with the [O{\small III}] emission and hence likely to be associated with the highly-ionized gas locally heated by intense radiation from the two clusters. The excess emission probably represents a dust feature, which is found to be spatially correlated with the OH absorption and a CO cloud. We find that a dust model including micron-sized graphite grains can reproduce the observed spectrum with the dust feature fairly well.}

\maketitle

\section{Introduction}
The center of our Galaxy exhibits very complicated structures of the interstellar medium (ISM) on the spatial scales of $\sim$1~pc to $\sim$100~pc, which harbors several active star formation sites (e.g., \citealt{Yus09}). For example, the most massive young stellar clusters in our Galaxy, the Arches and the Quintuplet cluster (\citealt{Cot92}; \citealt{Nag95}; \citealt{Ser98}; \citealt{Fig99}), are located near the Galactic center, which are thought to be responsible for the ionization of the two prominent filamentary structures, the Arched Filaments (\citealt{Cot96}; \citealt{Lan01}) and the Sickle (\citealt{Sim97}; \citealt{Fig99}; \citealt{Rod01}), respectively. Intense stellar winds and ultra-violet (UV) radiation from these massive clusters are likely to have given major impacts on their ambient ISM (\citealt{Lan05}). However the structures are so complicated that the clouds associated with and influenced by these clusters are rather difficult to clearly resolve. 

In the far-infrared (IR), the ISM in the Galactic center region has been studied by both gas forbidden line and dust continuum emission. Genzel et al. (1990) observed the strong far-IR fine-structure [C{\small II}] line at the Radio Arc and the +20/+50 km s$^{-1}$ molecular clouds with Kuiper Airborne Observatory, who suggested that the [C{\small II}] emission comes from interfaces between the molecular clouds and ionized zones. Infrared Space Observatory (ISO) detected various far-IR emission lines, which were interpreted as a result of excitation by the UV radiation produced by the Arches and the Quintuplet cluster (\citealt{Rod01}). Cotera et al. (2005), Simpson et al. (2007), and Yasuda et al. (2009) showed that the ratio of the highly-ionized to the lowly-ionized line intensity decreases with the distance from the Arches cluster with ISO, Spitzer, and AKARI, respectively, which indicates that the Arched Filaments are ionized by the Arches cluster. 


As for early studies on dust emission in the Galactic center, for example, Cox \& Laureijs (1989) estimated the averaged dust temperature to be 27~K from the far-IR observations of a central 450$\times$300~pc region with IRAS. Lis et al. (2001) derived the dust temperature of 18~K for the cold giant molecular cloud at $\ell\sim$0.2$^{\circ }$ to 0.6$^{\circ }$ and $b\sim\pm~0.1^{\circ }$ with ISO; they found that the contribution of warm dust ($\sim$35~K) was very small. Rodr\'\i guez-Fern$\acute {\rm a}$ndez et al. (2004) observed the 18 molecular clouds located far from thermal radio continuum sources in the Galactic center, showing that the dust continuum emission could be modeled with two temperatures: a cold component with temperature of $\sim$15~K and a warm component with temperatures varying between 25 and 40~K from source to source. Very recently, Etxaluze et al. (2011) analyzed the Herschel and ISO data of the Galactic center region, who found that the spectral energy distribution (SED) of central 2~pc in Sagittarius A$^{*}$ (Sgr A$^{*}$) could be fitted by a three-temperature modified blackbody curve with temperatures of 90, 45, and 23~K. Fitting a single-temperature modified blackbody curve to the PACS 70~$\mu$m and SPIRE 250, 350, and 500~$\mu$m data, they reported that dust temperatures for the Arched Filaments, the Sickle, and the Radio arc are $\sim$35~K, $\sim$35~K, and $\sim$28~K, respectively, while those for most other regions around the Galactic center are below $\sim$25~K. Hence these past studies consistently revealed that warm dust is a relatively minor constituent in the far-IR continuum emission, despite the presence of the very active star formation sites. In fact, Nakagawa et al. (1998) found that the Galactic center shows a global depression in [C{\small II}]/far-IR ratio along the Galactic plane; with the help of ISO [O{\small I}] data, Yasuda et al. (2008) concluded that the relatively strong far-IR continuum emission toward the Galactic center is caused mostly by non-C-ionizing soft radiation from old stellar populations. 

In the present paper, we report the result of the far-IR spectral mapping of the Arches and Quintuplet cluster region near the Galactic center, using the Fourier Transform Spectrometer (FTS) of the Far-IR Surveyor (FIS) onboard AKARI (Murakami et al. 2007; Kawada et al. 2007, 2008). AKARI FIS-FTS covers the wavenumber range of 60--140~cm$^{-1}$ with two kinds of two-dimensional detector arrays, and thereby we have simultaneously obtained the line and continuum maps of ionized sources and clouds around the clusters in spatially and spectrally continuous ways, which contain important information on the ISM in the Galactic center region. The early results for the line emission maps of the same region were already reported in Yasuda et al. (2009). In this paper, we present the spatial variations of spectral shapes mostly focusing on dust continuum emission. We report not only on the dust clouds associated with and influenced by the Arches and Quintuplet clusters as an original purpose of this study, but also about a dust feature, which is unexpectedly detected in spectra mainly toward the Arches cluster region.

\section{Observation and Data reduction}

\begin{figure}
   \centering
   \includegraphics[width=0.45\textwidth]{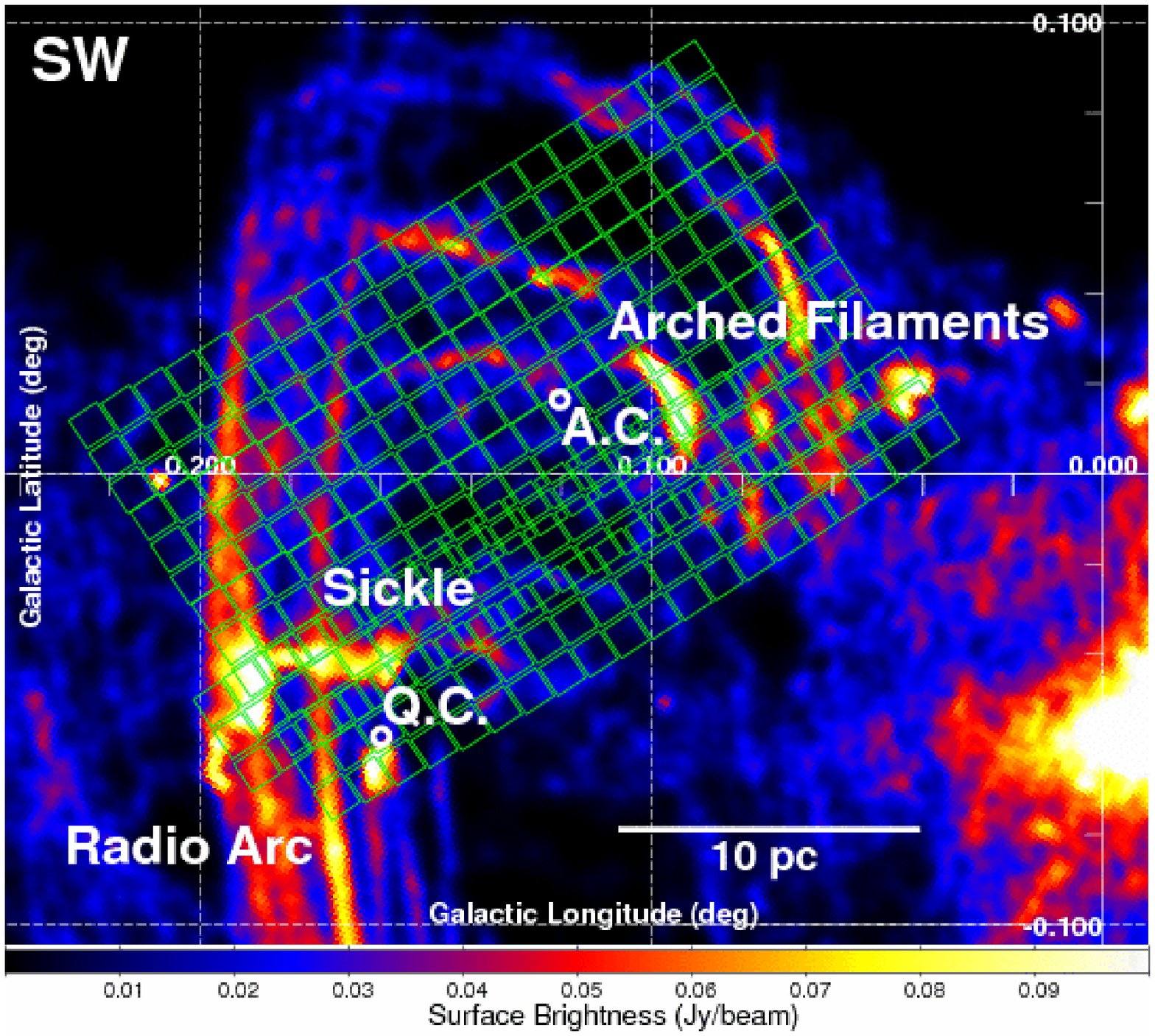}
   \includegraphics[width=0.45\textwidth]{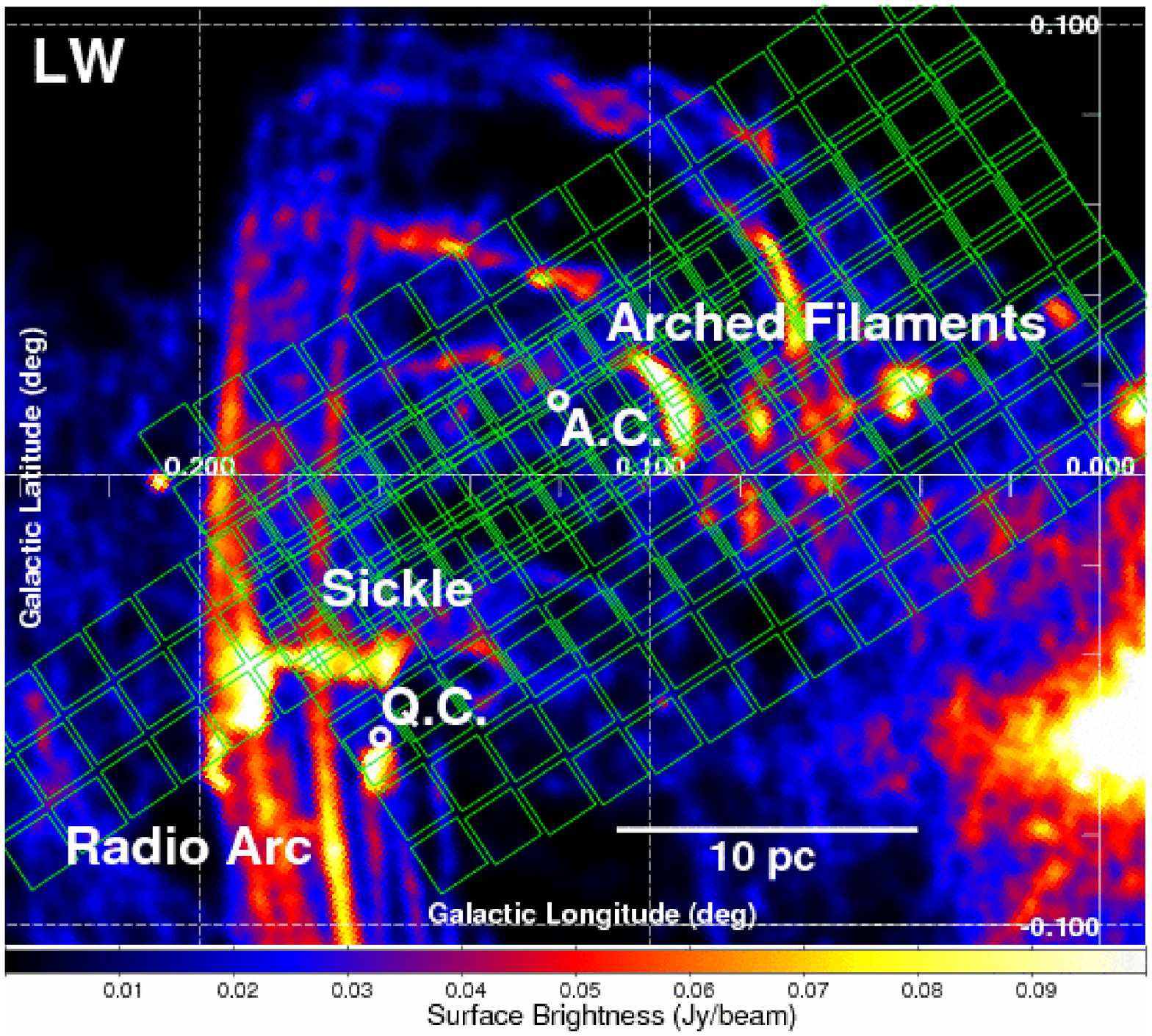}
   \caption{Areas mapped with the SW (top) and the LW (bottom) array, both overlaid on the radio 20~cm continuum map (\citealt{Yus84}). The positions of the Arches and Quintuplet clusters (here and hereafter, A.C. and Q.C. in the figures), together with the names of primary structures, are added in the panels. The field of view of each pixel is shown in the small box. }
   \label{fig1}
\end{figure}

\begin{figure*}
   \centering
   \includegraphics[angle=-90, width=0.45\textwidth]{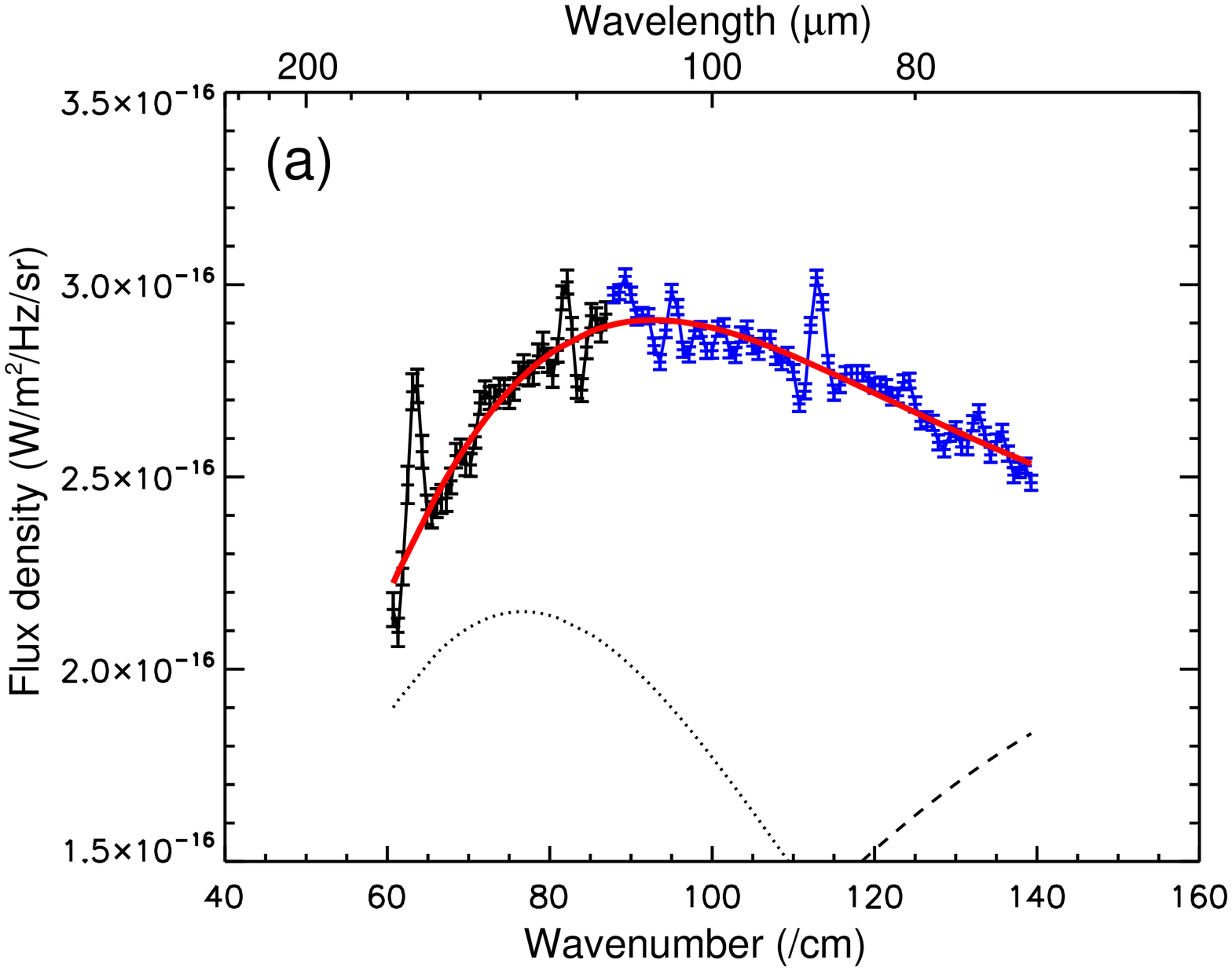}\includegraphics[angle=-90, width=0.45\textwidth]{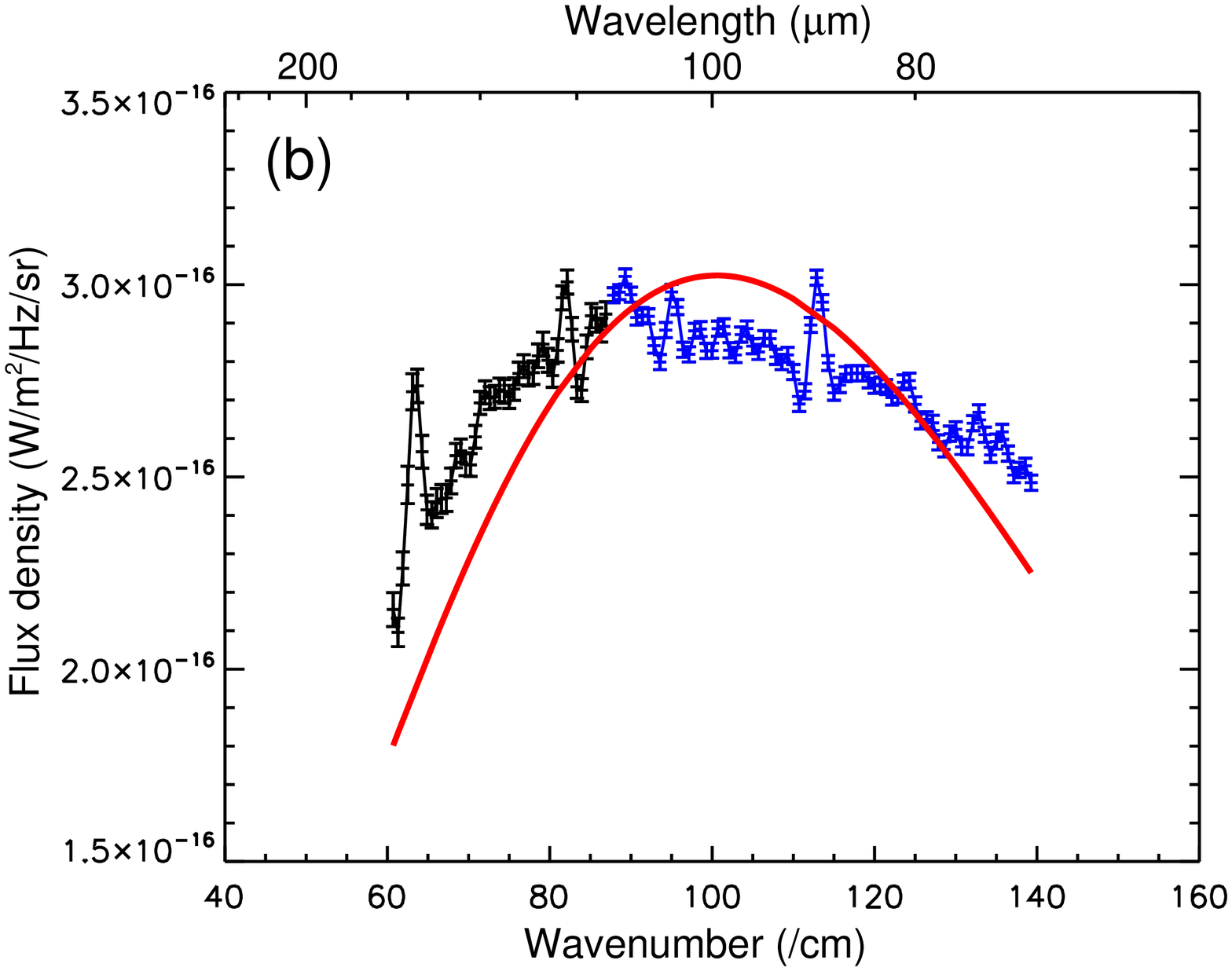}\\
   \includegraphics[angle=-90, width=0.45\textwidth]{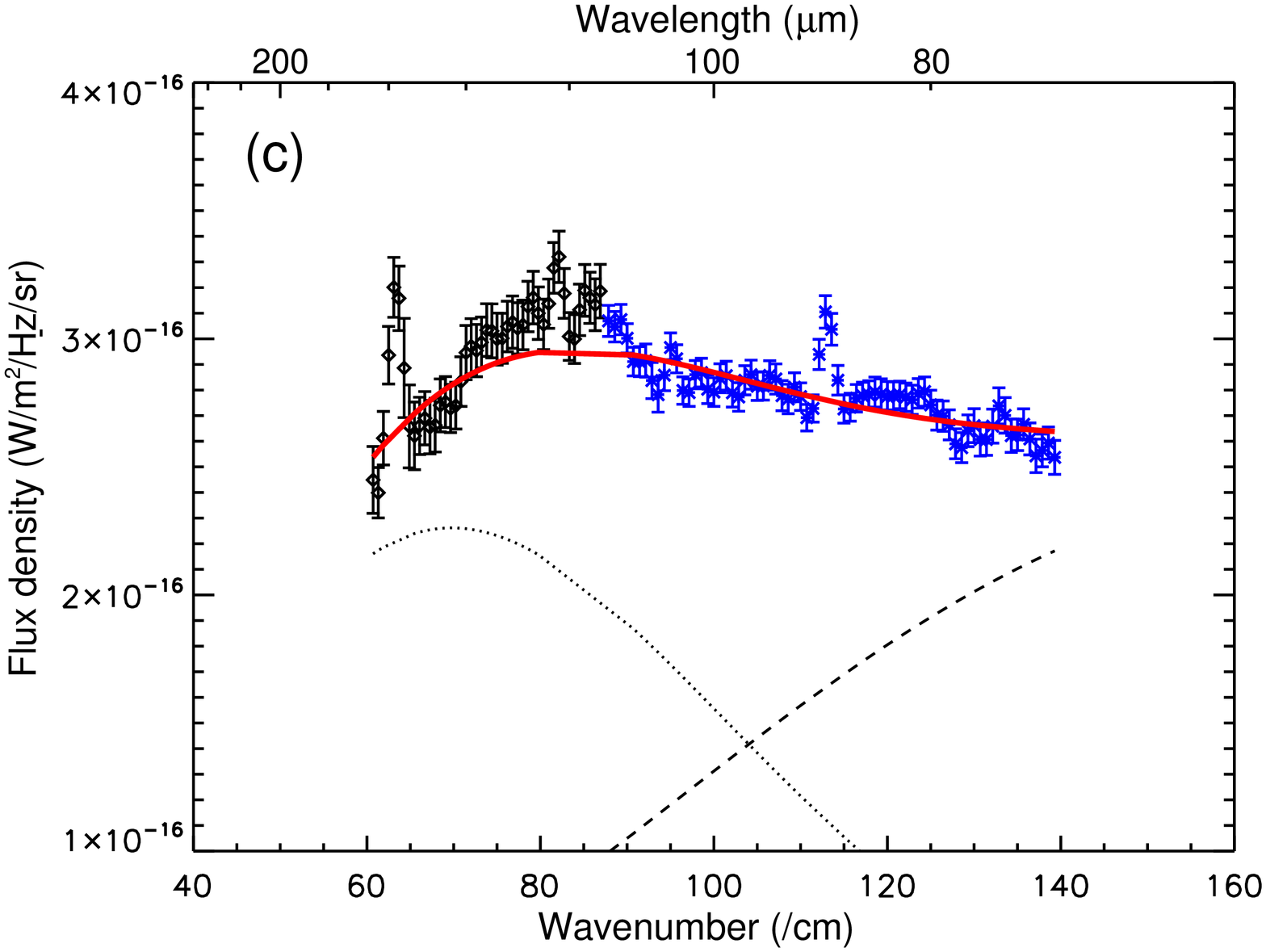}\includegraphics[angle=-90, width=0.45\textwidth]{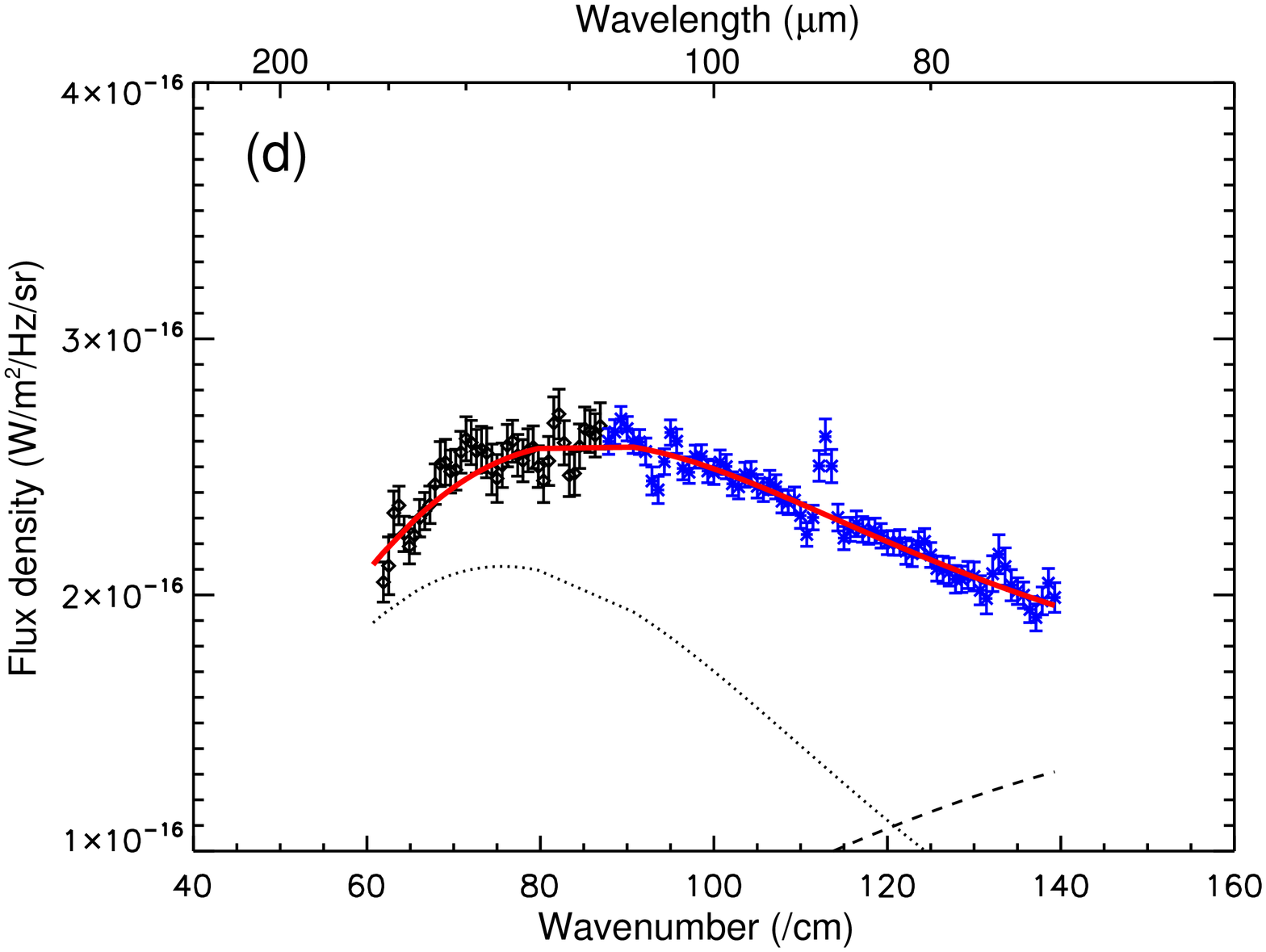}
   \caption{AKARI/FIS-FTS spectra of (a, b) the total region covered by both SW and LW, (c) the regions toward the Arches cluster and (d) the Quintuplet cluster. The results of the double-temperature (panels a, c, d) and single-temperature (panel b) modified blackbody model fitting are shown together, where the wavenumber ranges corresponding to the emission lines ([O{\small III}]: 111 to 115~cm$^{-1}$, [N{\small II}]: 81 to 85~cm$^{-1}$, [C{\small II}]: 62 to 66~cm$^{-1}$) and the excess feature (80 to 90~cm$^{-1}$) are masked in the spectral fitting. Here and hereafter the different colors are used to show the spectra derived from the SW (blue) and LW (black) detectors. The aperture regions used to extract spectra (c) and (d) are indicated in Fig.~\ref{fig4}a}
   \label{sed_all}
\end{figure*}

The observations were carried out by 5 pointed observations in September 2006 and March 2007. AKARI FIS-FTS has two kinds of detector arrays (WIDE-S and WIDE-L; hereafter SW and LW, respectively), and we obtain the spectra of the SW (88--140~cm$^{-1}$, 71--114~$\mu$m) and the LW range (60--88~cm$^{-1}$, 114--167~$\mu$m), simultaneously. The effective angular resolution is $39''-44''$ for SW and $53''-57''$ for LW (\citealt{Kaw08}). The full-resolution mode was used with a very short reset interval of 0.1~sec. Figure~\ref{fig1} shows the areas mapped with the SW and LW arrays, where the radio 20 cm continuum map (\citealt{Yus84}) is superposed. The difference in the spatial coverage between the SW and the LW map is due to the difference in the field of view between the SW and the LW array detector (\citealt{Kaw08}). As can be seen in the figure, both SW and LW cover the Arches and Quintuplet clusters together with the Arched Filaments and the Sickle.  

We used the official pipeline for data analyses. Details of the data reduction processes and calibration were described in Murakami et al. (2010). Although the observational data were obtained in the full-resolution mode ($R=0.19$~cm$^{-1}$), we used an SED mode ($R=1.2$~cm$^{-1}$) analysis to increase the signal to noise ratios for continuum emission, because the main purpose of this paper is to study dust emission. We used both short sides of the FTS interferogram with respect to the center burst for the SED analysis rather than one long side for the full-resolution analysis; due to the limitation in the movement of the FTS scanning mirror, only one side is set to be long for the optical path difference in the full-resolution mode (\citealt{Kaw08}). The usage of both sides increases tolerance against the effects of detector artifacts such as transient response. Another merit of the SED mode analysis is that it produces spectra much less affected by the channel fringes than the full-resolution mode analysis. 

It should be noted that the pixels of the SW and LW detectors do not observe the same sky. Therefore, we first created a map for each spectral bin of 0.6~cm$^{-1}$ for both SW and LW and then regridded it with the common spatial bin of $30''\times30''$. At this stage, we derived 122 maps with a spectral bin of 0.6~cm$^{-1}$ to create a spectrum at every spatial bin of $30''\times30''$ for the region overlapped between the SW and LW maps. Then using fluxes in two contiguous spectral bins and eight contiguous spatial pixels, we averaged 27 (=$3~\times~9$) fluxes in total to obtain a local spectrum for every spectral bin. Finally the spectral mismatches, caused mainly by the difference in field-of-view and also possibly by residual calibration uncertainties, were removed by scaling LW to match SW spectra at several overlapping and adjacent spectral bins by factors of 0.85--1.11.

Figure~\ref{sed_all}a shows the spectrum integrated over the total area covered by both SW and LW (Fig.~\ref{fig1}). The SW and LW spectra are combined into one spectrum as described above.
From the total spectrum and local spectra, we detect the three far-IR fine-structure lines, [O{\small III}] 88~$\mu$m (113~cm$^{-1}$), [N{\small II}] 122~$\mu$m (82~cm$^{-1}$), and [C{\small II}] 158~$\mu$m (63~cm$^{-1}$) as well as the OH absorption at 119~$\mu$m (84~cm$^{-1}$) due to the fundamental rotational mode, and dust continuum emission over the whole wavenumber range of FIS-FTS.

\begin{figure*}
   \centering
   \includegraphics[width=0.3\textwidth, angle=-0]{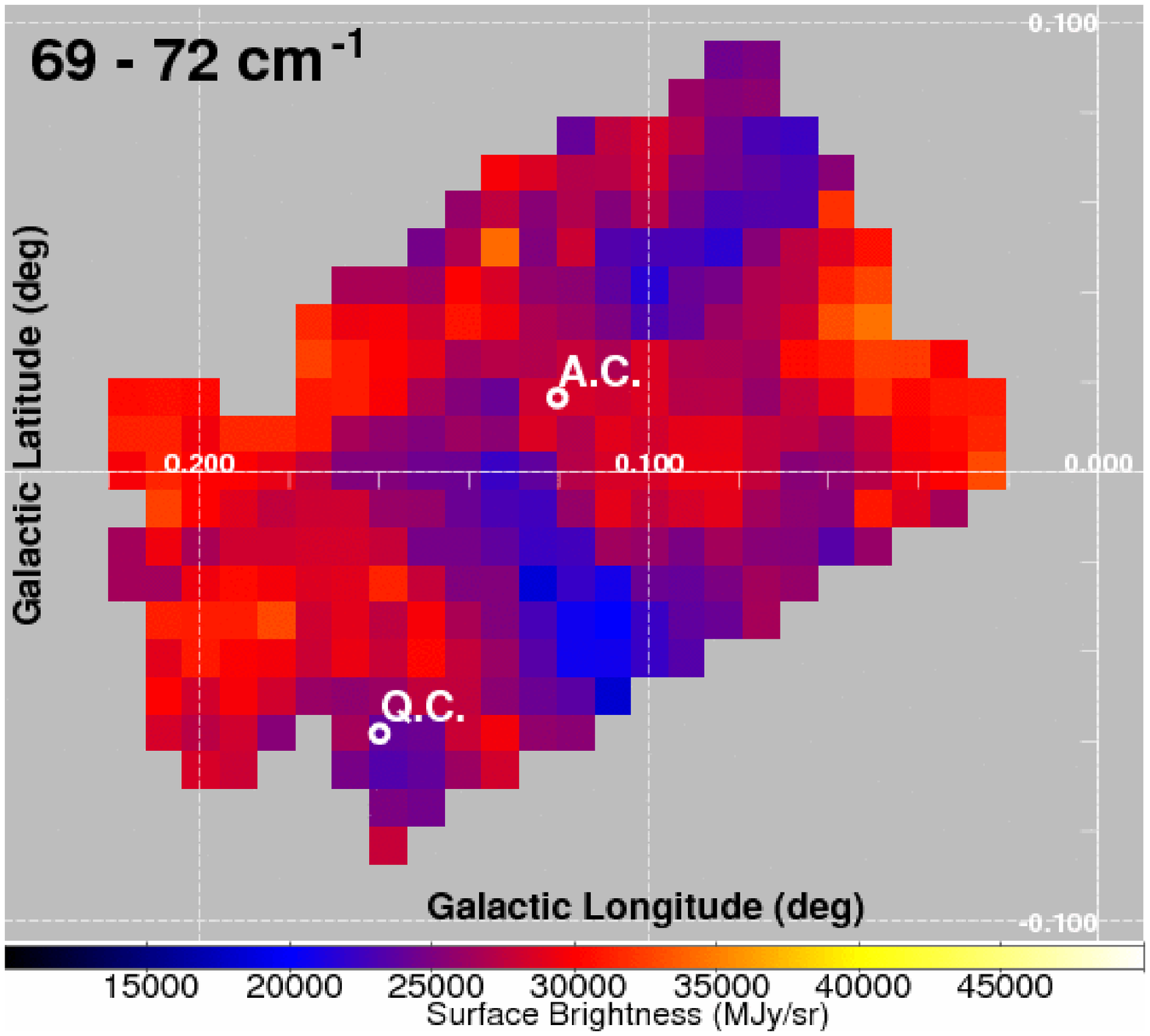}
   \includegraphics[width=0.3\textwidth, angle=-0]{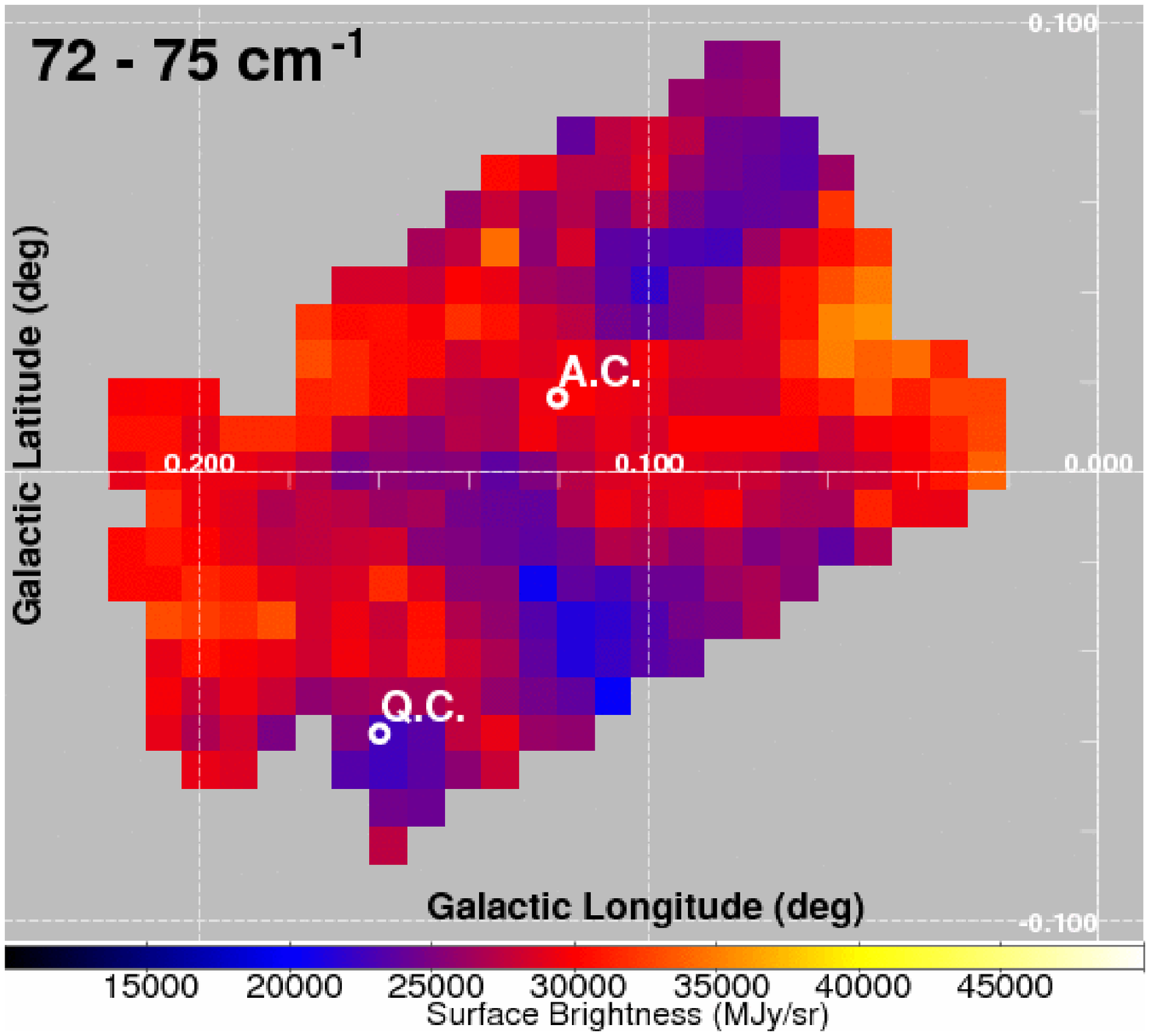}
   \includegraphics[width=0.3\textwidth, angle=-0]{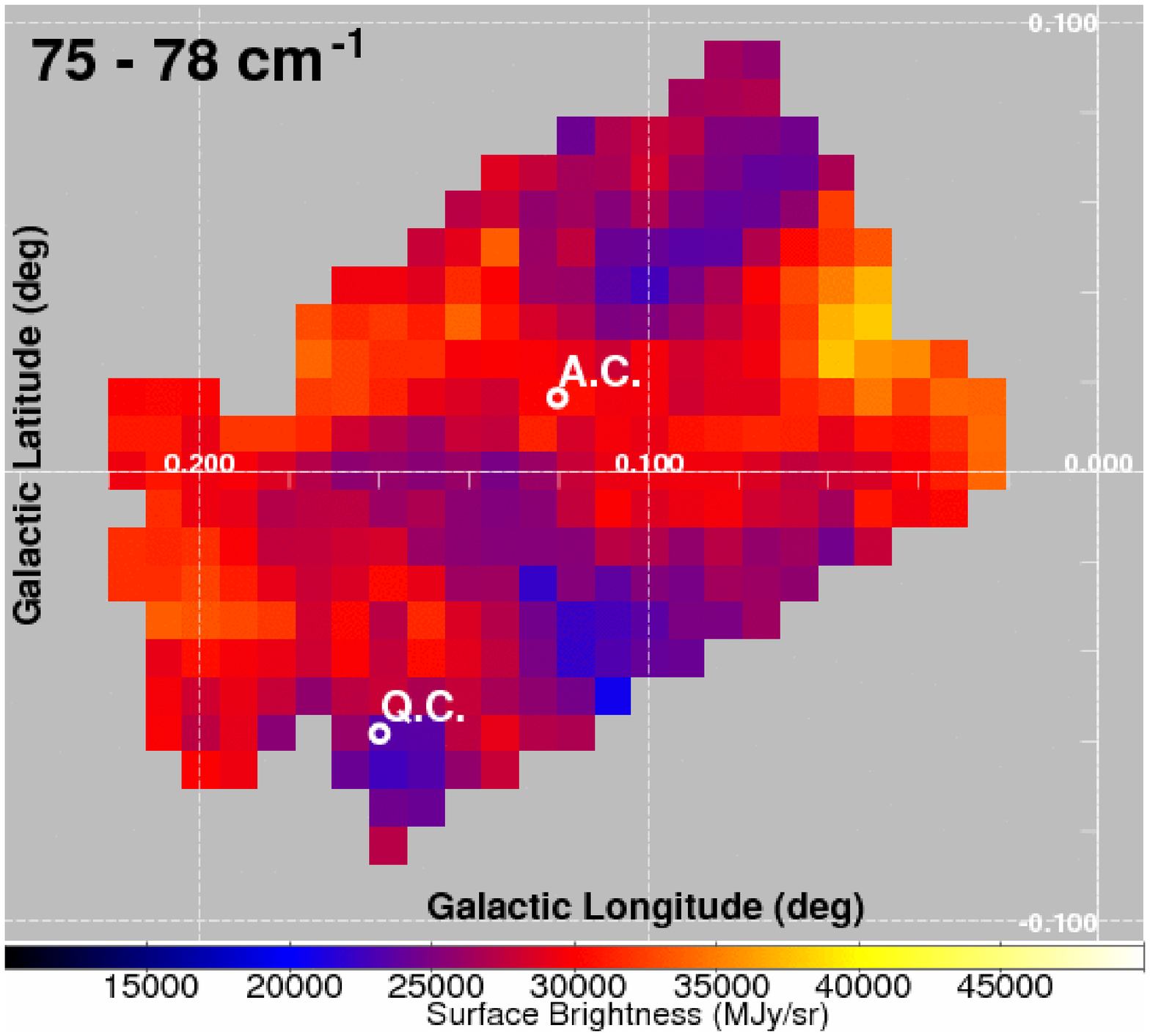}
   \includegraphics[width=0.3\textwidth, angle=-0]{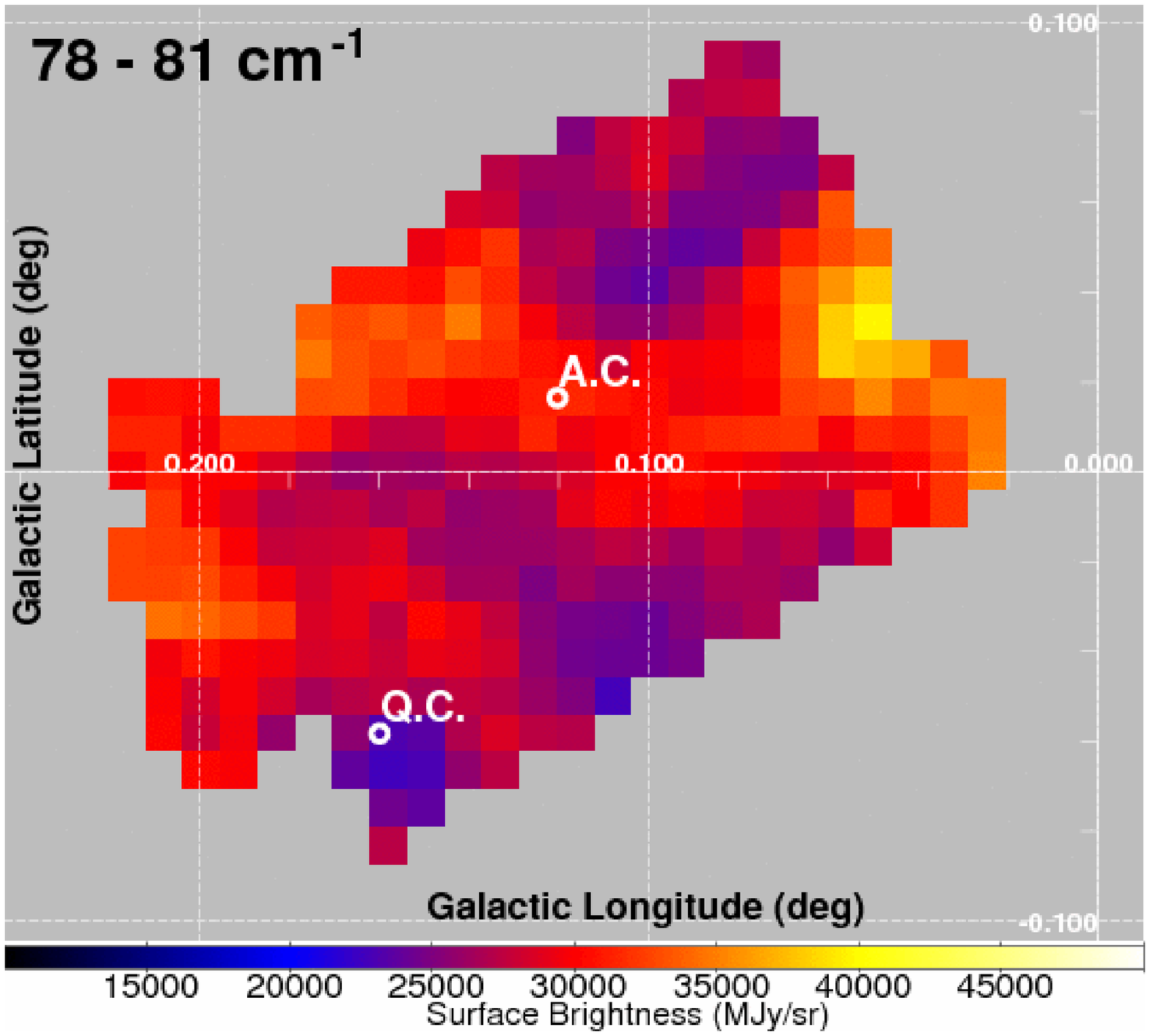}
   \includegraphics[width=0.3\textwidth, angle=-0]{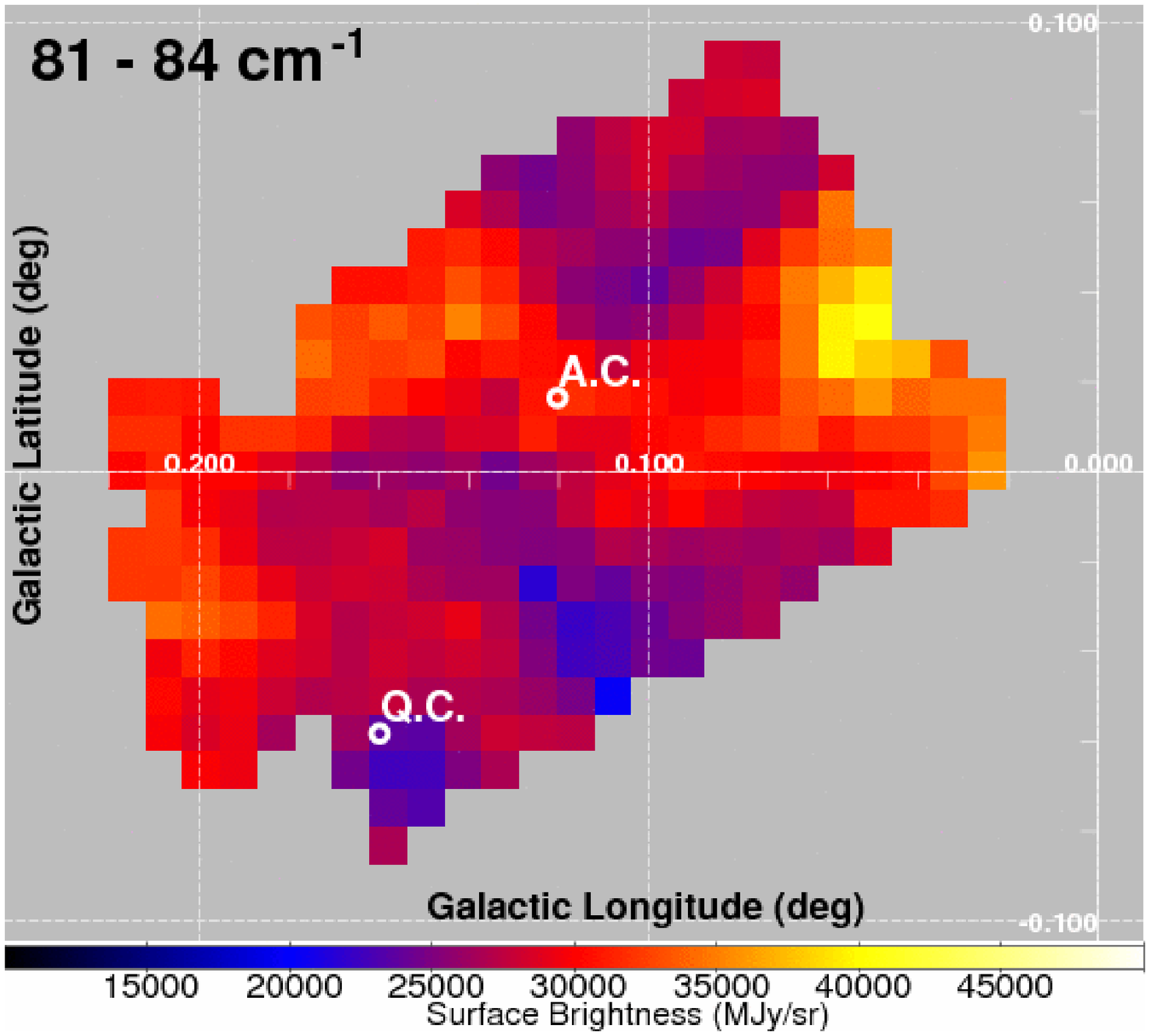}
   \includegraphics[width=0.3\textwidth, angle=-0]{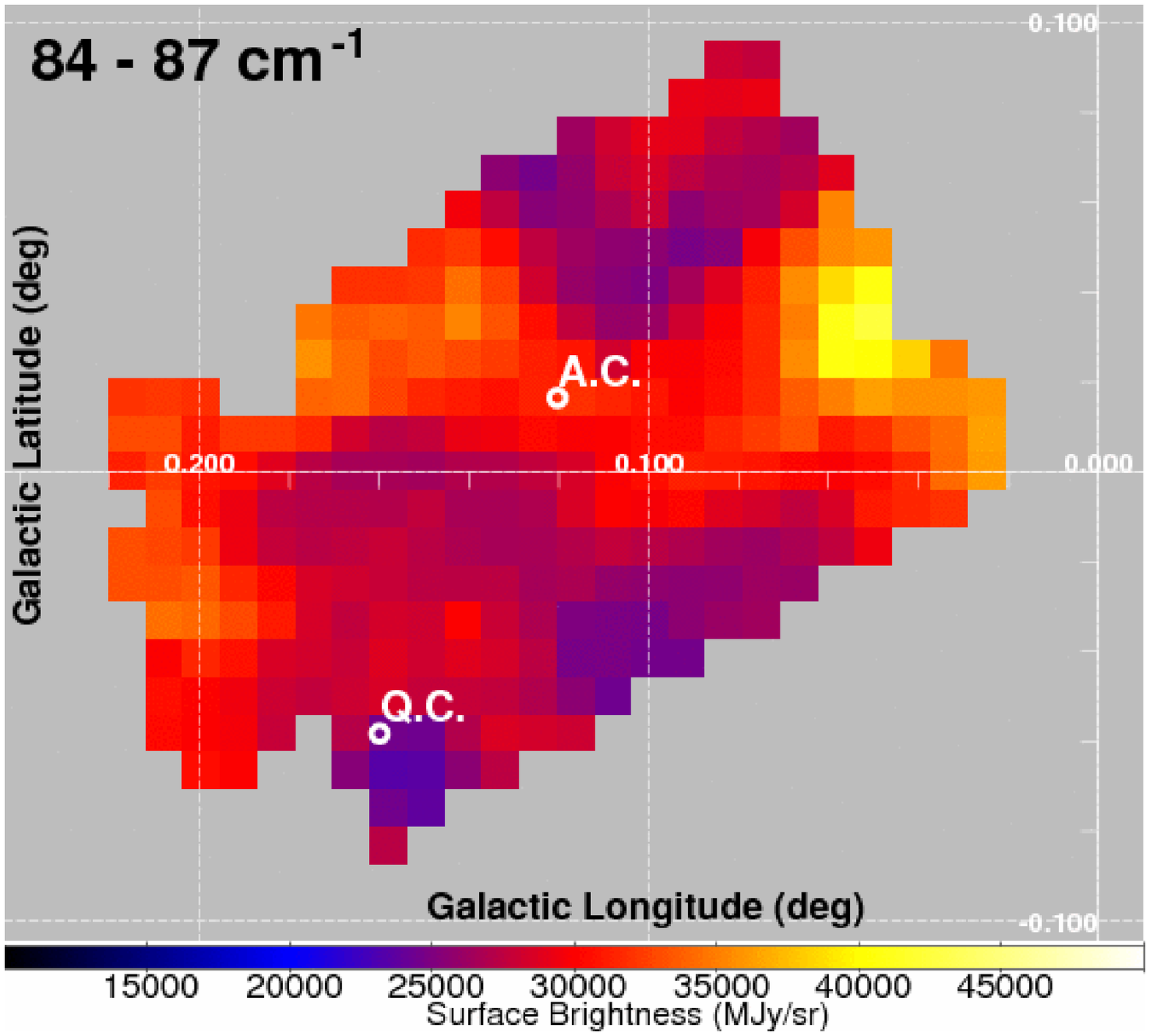}
   \includegraphics[width=0.3\textwidth, angle=-0]{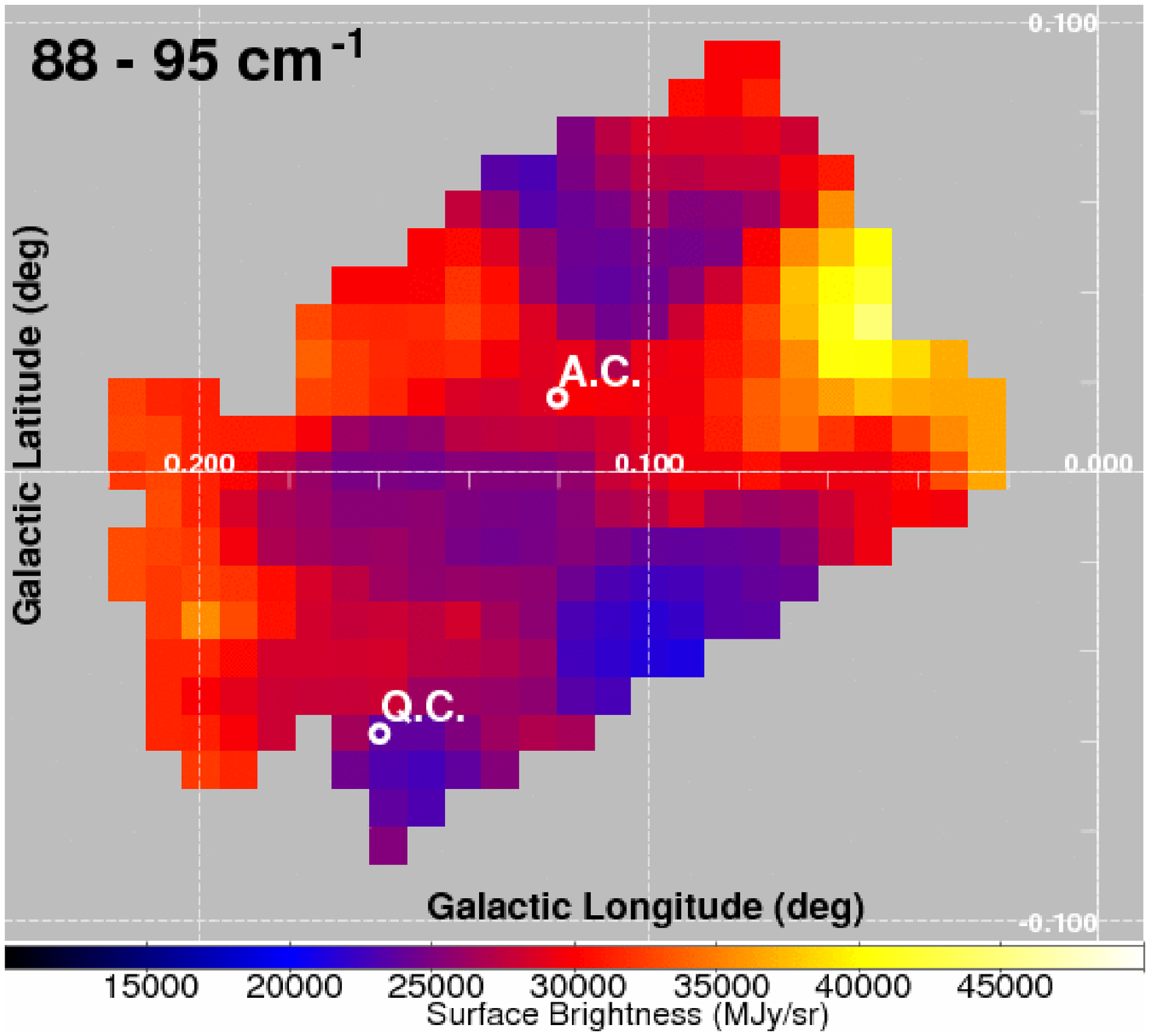}
   \includegraphics[width=0.3\textwidth, angle=-0]{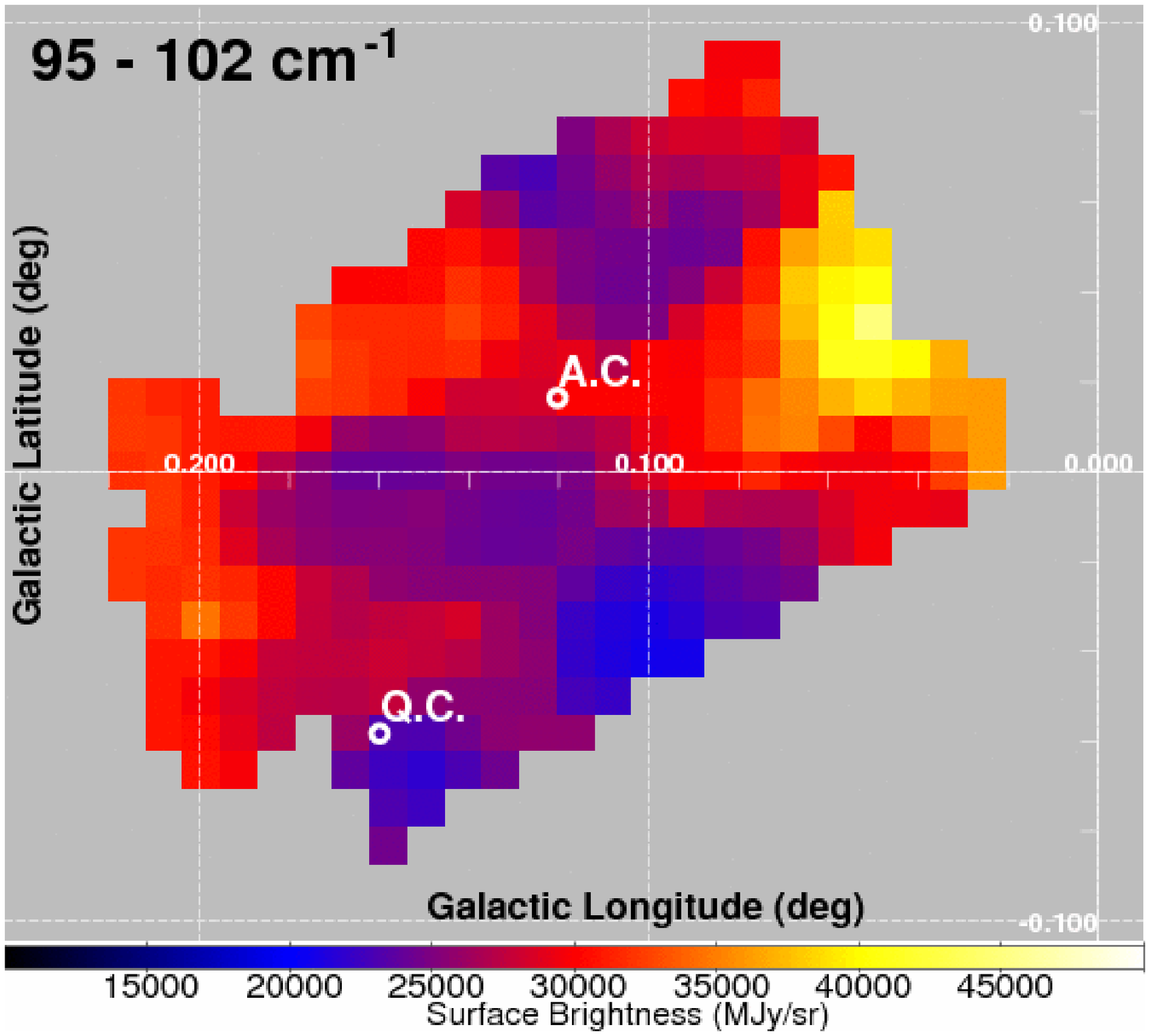}
   \includegraphics[width=0.3\textwidth, angle=-0]{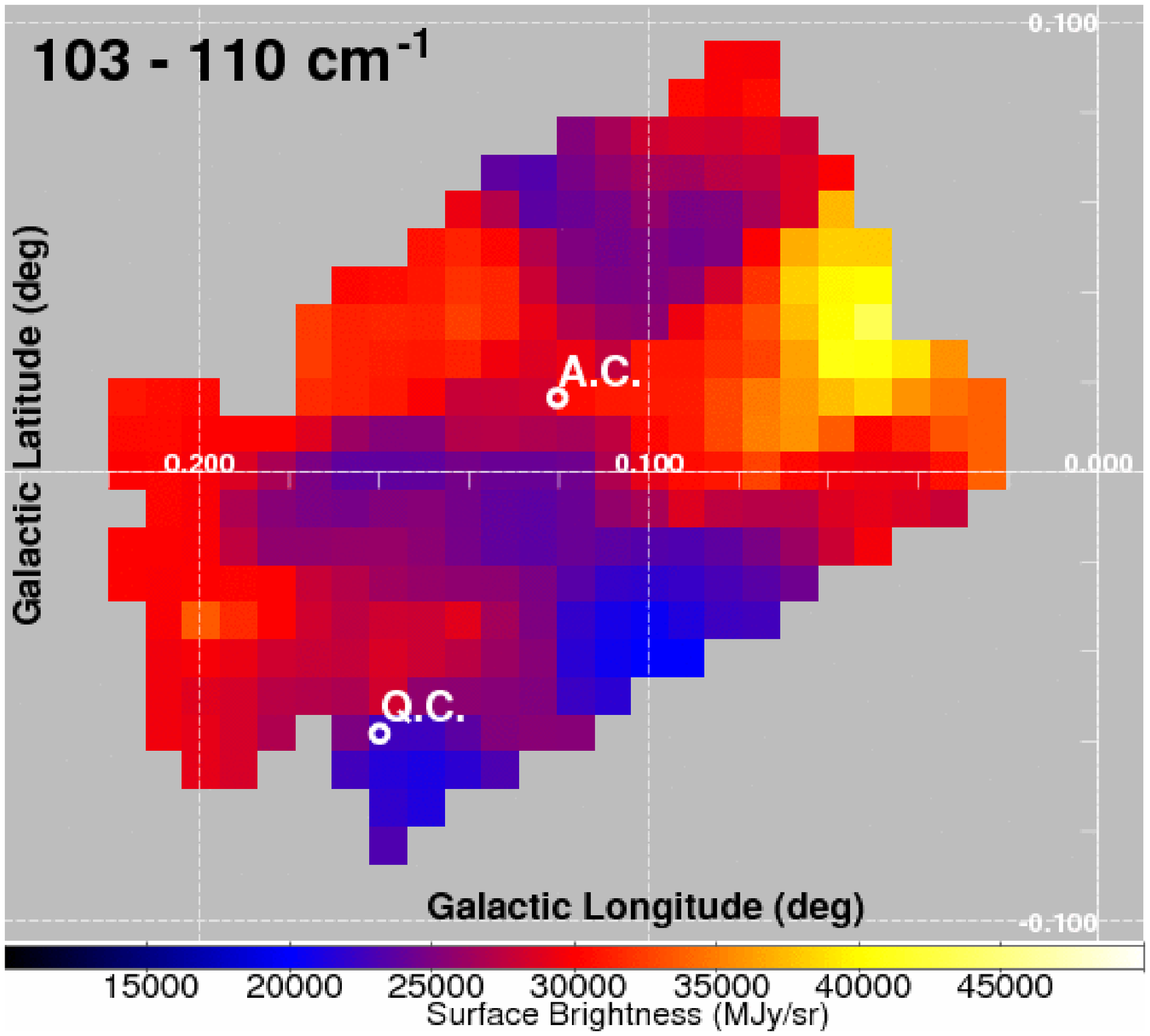}
   \includegraphics[width=0.3\textwidth, angle=-0]{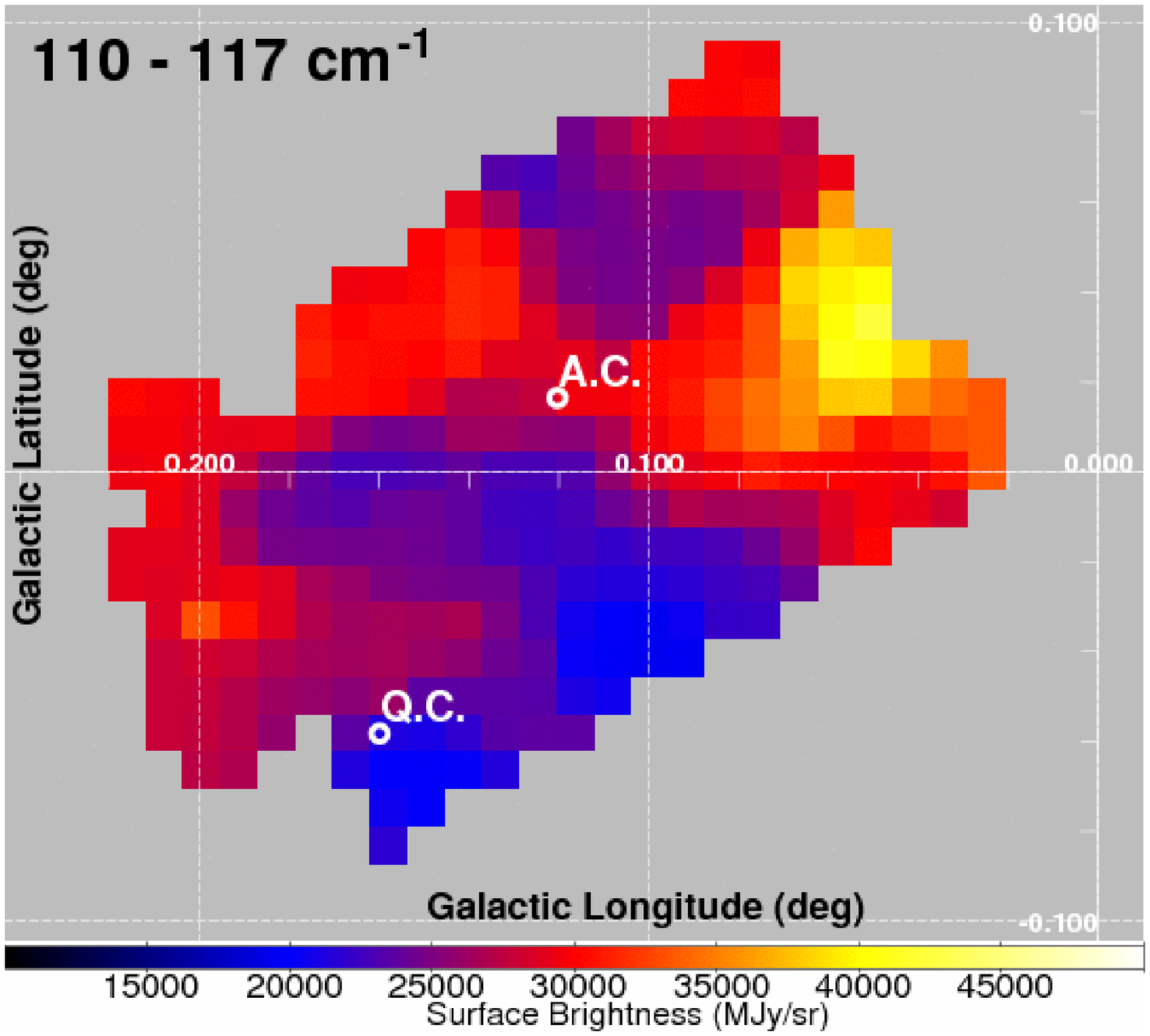}
   \includegraphics[width=0.3\textwidth, angle=-0]{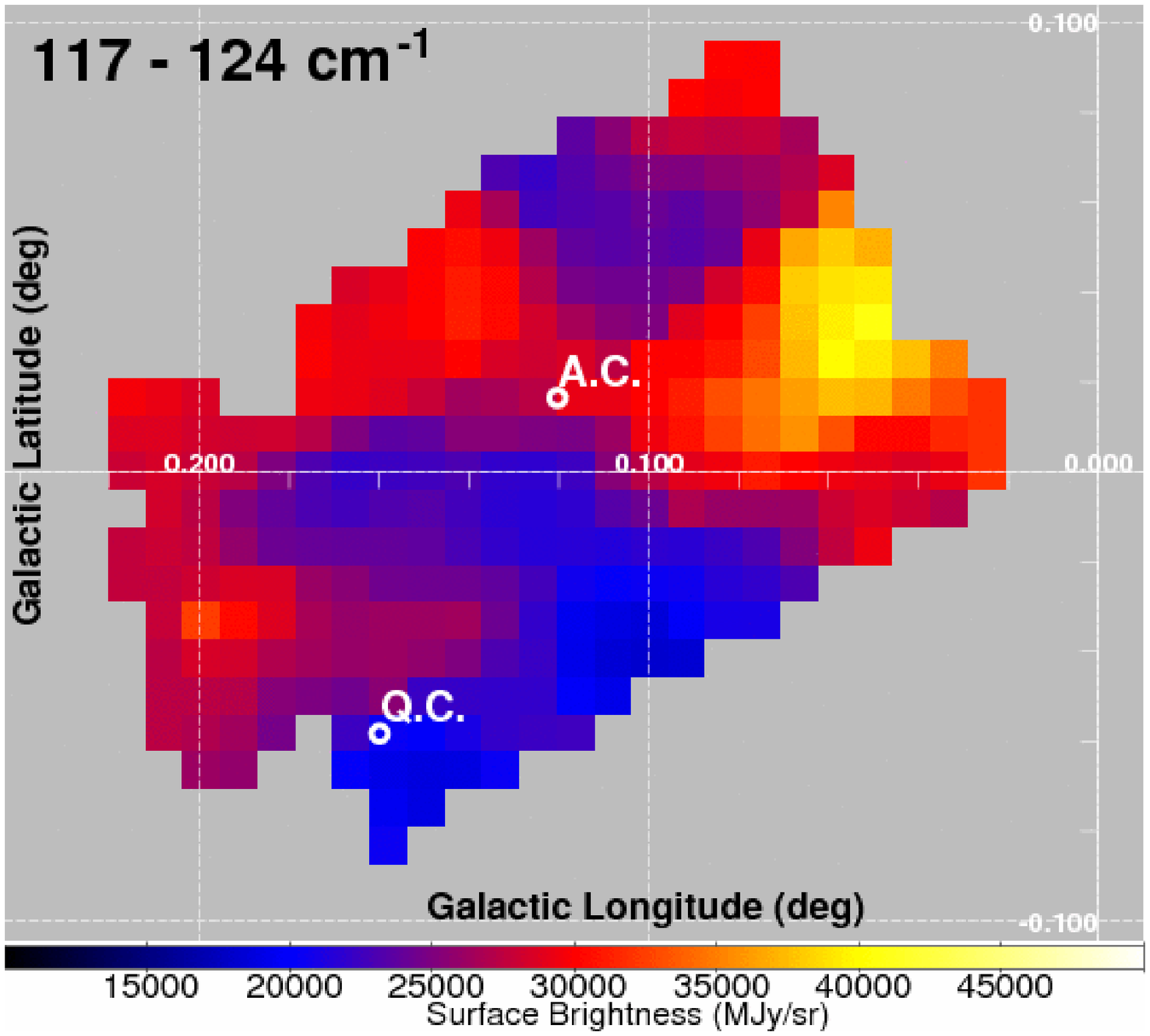}
   \includegraphics[width=0.3\textwidth, angle=-0]{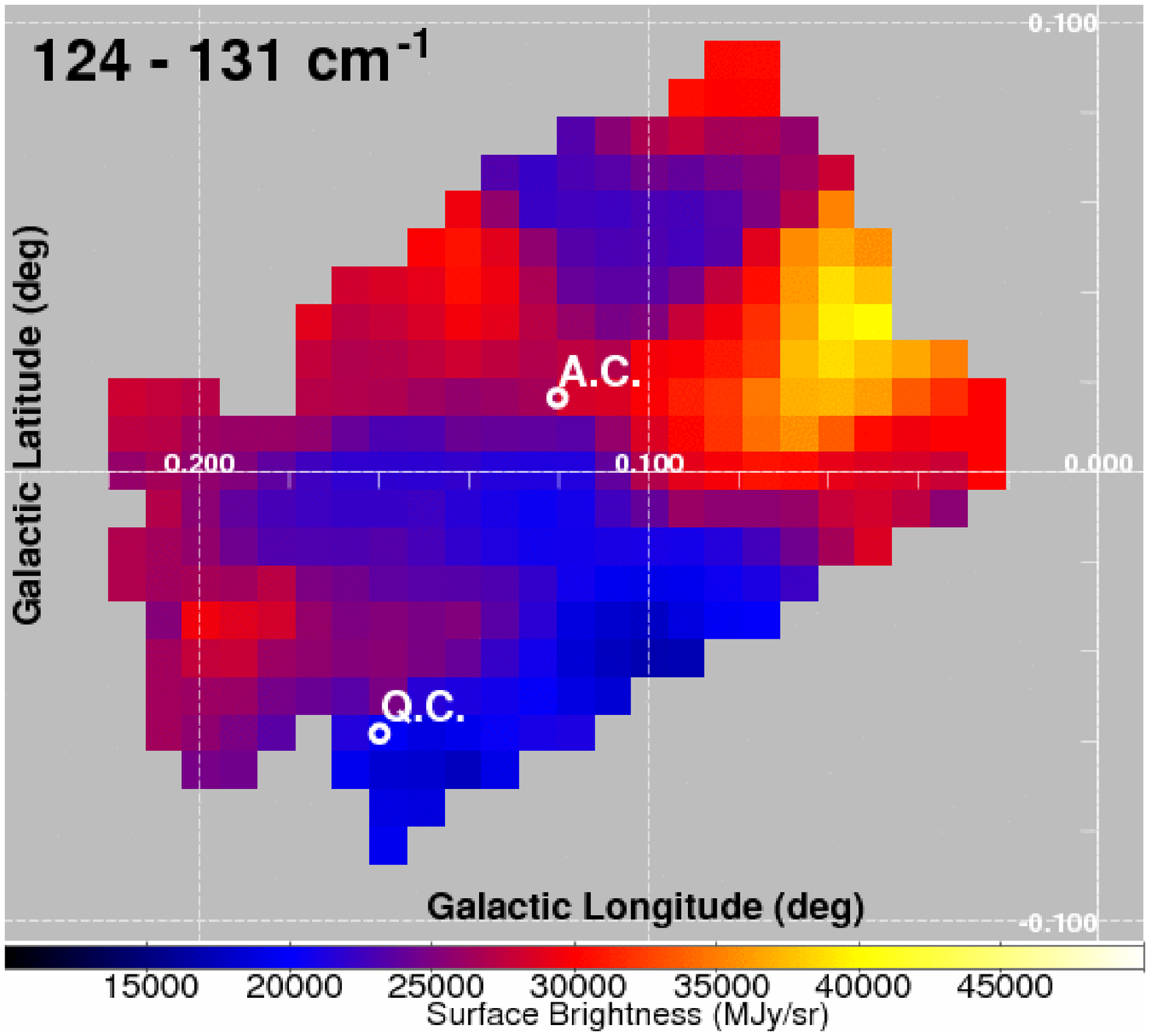}
   \caption{Far-IR continuum maps obtained by the LW (upper six) and SW (lower six maps) arrays of AKARI FIS-FTS for the denoted integration ranges of wavenumbers. The surface brightness levels are given on common color scales among the maps.}
   \label{fig3}
\end{figure*}

\section{Results}
\subsection{Dust continuum maps}
We derived continuum maps, using the local spectrum per spectral bin, after masking the spectral line regions, [O{\small III}] for the SW spectrum and [N{\small II}] and [C{\small II}] for the LW spectrum. 
Figure~\ref{fig3} shows the far-IR continuum maps thus obtained for the different integration ranges of wavenumbers. The figures show similar distributions of the far-IR continuum emission among the maps at different wavenumbers, following the arch structure composed by the Radio Arc, the Arched Filaments, and their bridge. The similarity suggests that the dust temperature does not vary much over the observed region. 
The local peaks of the SW and LW continuum maps near the Galactic center agree with each other, which are located near the H5 H{\small II} region at ($\ell,b$)~$\simeq$~(0$^{\circ }$.04, 0$^{\circ }$.02) (\citealt{Zha93}; \citealt{Lan10}). 

\begin{figure}
   \centering
   \includegraphics[width=0.45\textwidth]{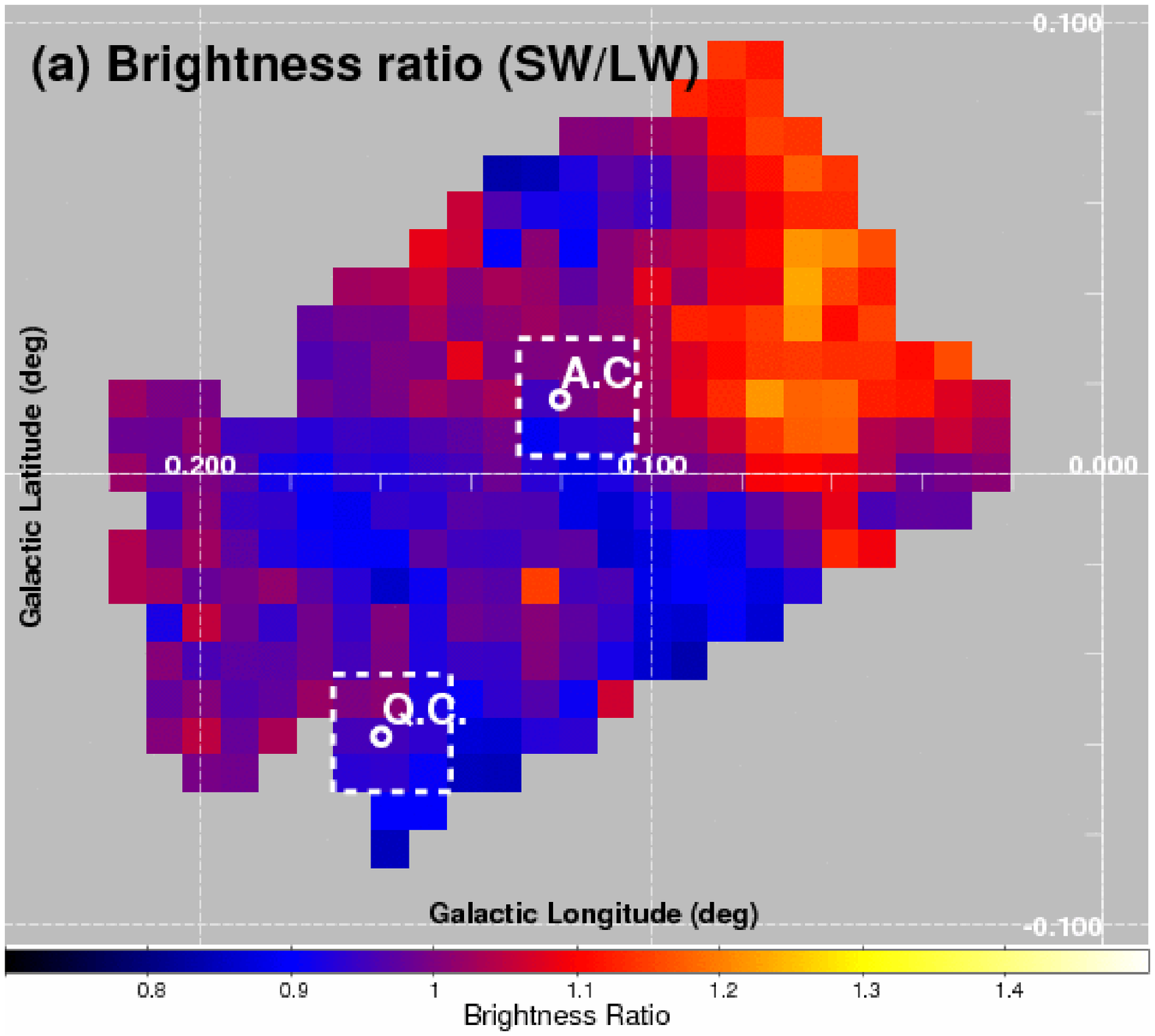}
   \includegraphics[width=0.45\textwidth]{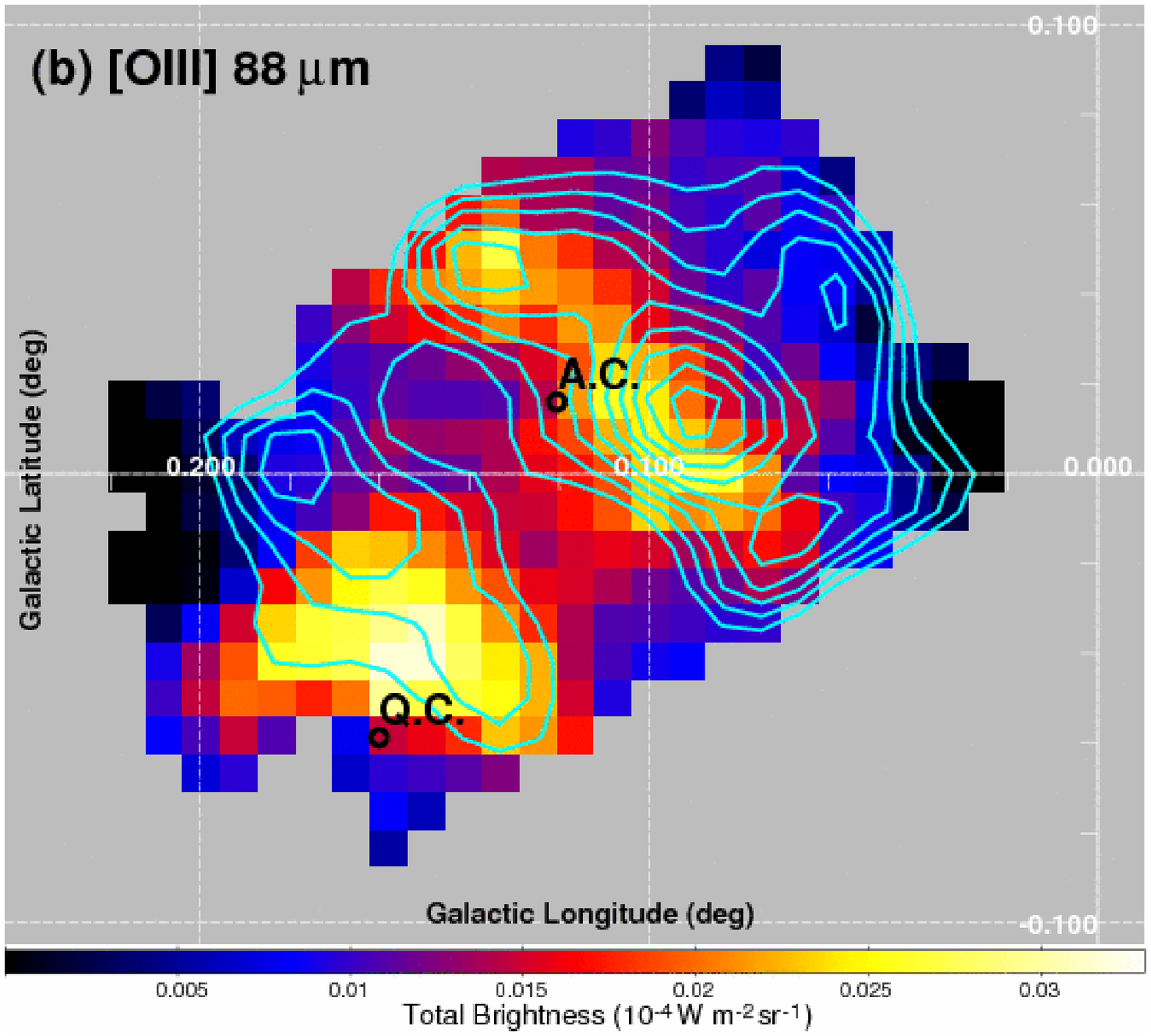}
      \caption{(a) Brightness ratio map derived from the SW continuum averaged over the range of 88--131~cm$^{-1}$ that is divided by the LW continuum over 66--87~cm$^{-1}$. The dashed squares indicate the aperture regions used to obtain the spectra shown in Figs.~\ref{sed_all}c and d. (b) [O{\small III}] 88~$\mu$m line intensity map overlaid on the contours of the [N{\small II}] 122~$\mu$m line intensity. The contour levels are linearly scaled from 4$\times$10$^{-7}$ to 1.5$\times$10$^{-6}$~W~m$^{-2}$~sr$^{-1}$.}
      \label{fig4}
\end{figure}

We calculated the brightness ratio of the SW continuum averaged over the range of 88--131~cm$^{-1}$ to the LW continuum over 66--87~cm$^{-1}$, the spatial distribution of which is shown in Fig.~\ref{fig4}a. The map shows notably small variations in the brightness ratio from $\sim$0.9 to $\sim$1.2, which correspond to the color temperature range of $\sim$25 to $\sim$30~K by adopting a dust modified blackbody model with the emissivity power-law index of 2.
Thus the spatial variation of the color temperature is notably small, despite the fact that the mapped area contains both active clusters and cold molecular clouds. In particular, there is no appreciable increase in the color temperature near the Arches and the Quintuplet cluster. 

Figure~\ref{fig4}b shows the map of the [O{\small III}] line emission, where the contour map of the [N{\small II}] 122 $\mu$m line emission is overlaid. They are both obtained by the spectral fitting described below. The [O{\small III}] line map represents the distribution of highly-ionized gas, since the ionization potential of O$^{+}$ is as high as 35.1 eV. Judging from their spatial correspondence, the ionized gases are likely to be associated with the Arches and the Quintuplet cluster. The [N{\small II}] line emission is extended farther from the clusters than the [O{\small III}] line emission. Since the ionization potential of N is 14.5 eV, Fig.~\ref{fig4}b indicates that the ionization degree of the gases heated by the two clusters decreases considerably with the distance, which is consistent with the conclusions drawn by the past studies (\citealt{Cot05}; \citealt{Sim07}; \citealt{Yas09}). Thus, in the line emission, we clearly recognize that the ISM is influenced by the two clusters, which is in marked contrast to the brightness ratio of the dust continuum emission (Fig.~\ref{fig4}a).

\subsection{Spectra of various local areas}   

Figures~\ref{sed_all}c and d show the spectra obtained for the regions toward the Arches and the Quintuplet cluster, respectively. We adopt the $1'.5\times1'.5$ (three by three bins) apertures centered at the clusters. Most interestingly, the spectrum for the Arches cluster region appears to have a broad ($\sim15$~cm$^{-1}$) hump around 80~cm$^{-1}$. Besides, the spectrum for the Arches cluster region has a slightly bluer continuum than that for the Quintuplet cluster region at large wavenumbers, which is consistent with the brightness ratio map in Fig.~\ref{fig4}a. The [O{\small III}] line intensities are similar between the two, while the [N{\small II}] and [C{\small II}] lines in the spectrum for the Quintuplet cluster is weaker than those in the spectrum for the Arches cluster.

In order to fit the entire baseline of the total spectrum in Fig.~\ref{sed_all}a, we apply a two-temperature dust modified blackbody model, using an emissivity power-law index $\beta=2$ \cite{Pie00, Rod04} and masking the spectral line regions. As a result, we find that the two-temperature modified blackbody model reproduces the continuum shape very well and another temperature component is not required. It is obvious that a single-temperature model cannot reproduce the spectrum at all (Fig.~\ref{sed_all}b). The best-fit temperature of the warmer dust component is 51~K, while that of the colder dust component is 23~K (table~\ref{table1}); they roughly agree with the temperatures derived by Etxaluze et al. (2011). Then we use the same model to fit the spectra of the Arches and Quintuplet clusters, adding another masking region at 80 to 90~cm$^{-1}$ because of the apparent presence of the broad hump over these wavenumbers. Since the temperature of the warm dust component is not well constrained, we fix its temperature at 51~K, which is the best-fit value for the total spectrum. As a result, Fig.~\ref{sed_all} clearly shows that the spectrum toward the Arches cluster possesses a broad excess component on top of the dust modified blackbody continuum, which is too narrow to be explained by another modified blackbody component. In contrast, the spectrum toward the Quintuplet cluster does not exhibit such an excess component.

\begin{figure*}
   \centering
   \includegraphics[width=0.9\textwidth]{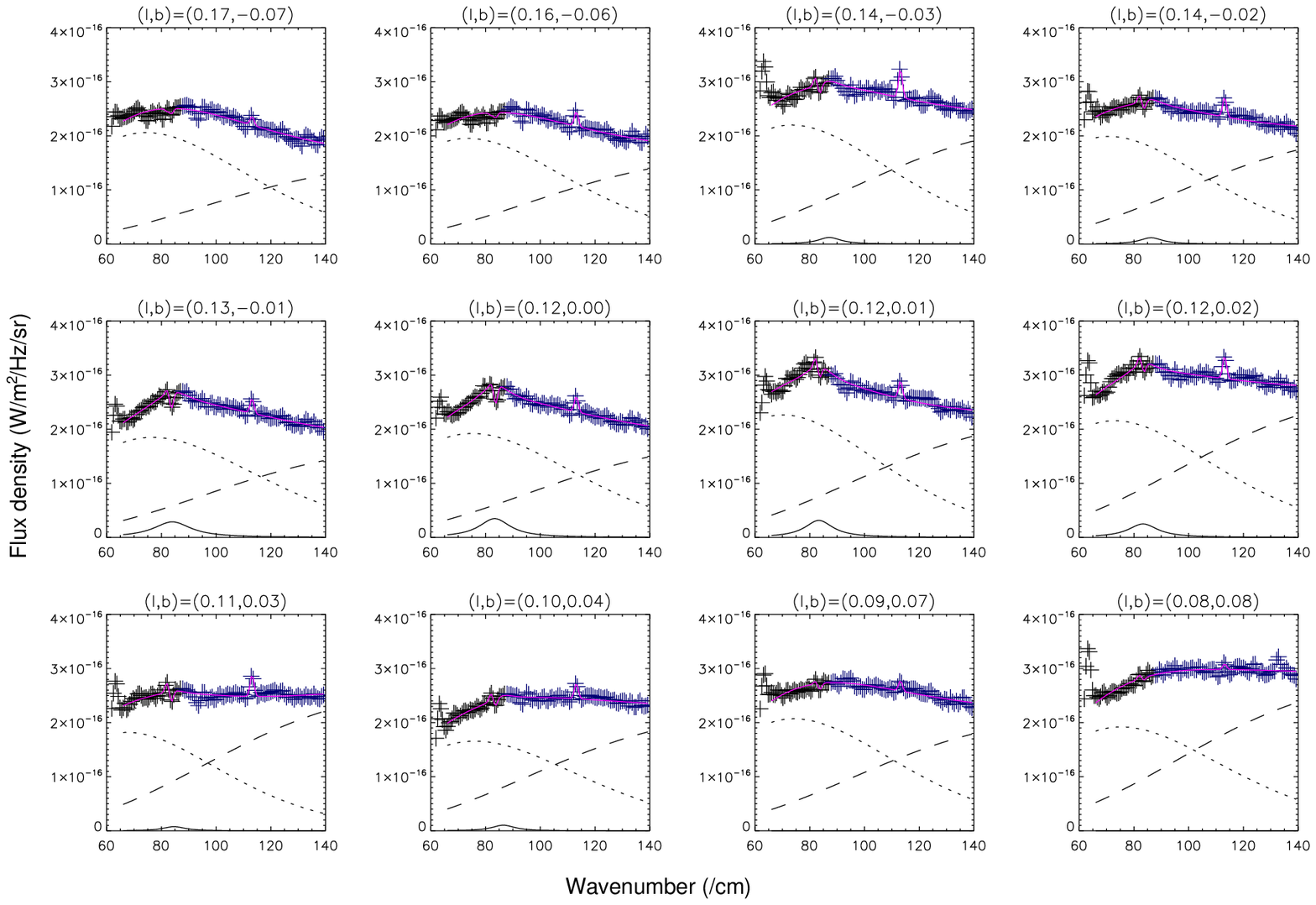}

\vspace{0.5cm}

   \includegraphics[width=0.9\textwidth]{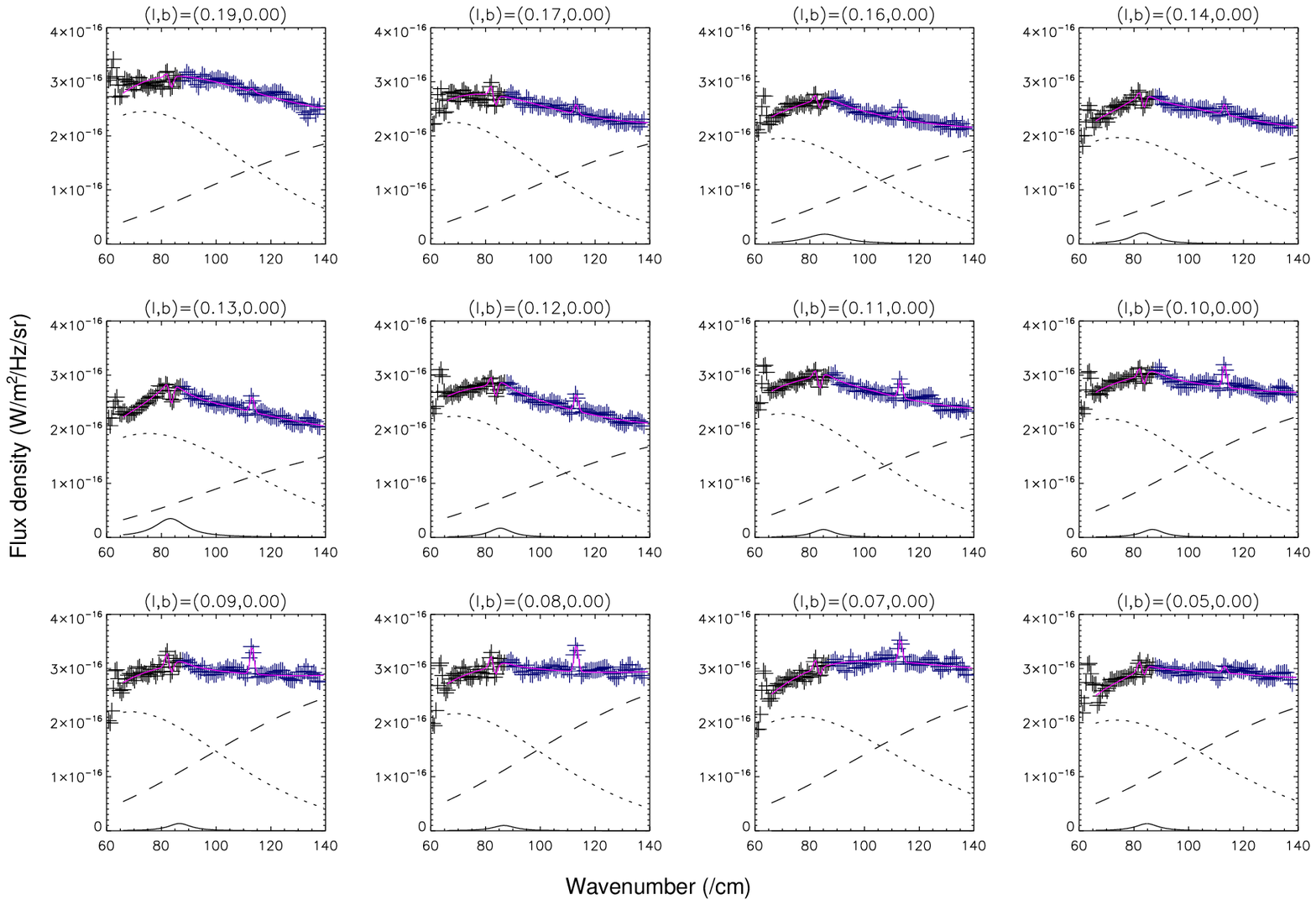}
      \caption{Spectra of local ($30''\times30''$) areas. The upper panels show the variation along the line connecting the Arches and Quintuplet clusters, while the lower panels show that along the Galactic plane. For each spectrum, its position is denoted at the top in the Galactic coordinates. The solid curves indicate the best-fit model that includes two dust modified blackbody components, the [O{\small III}] and [N{\small II}] emission lines, the OH absorption line, and the excess component represented by a Lorentian function (see text for details).}
         \label{fit_ex}
\end{figure*}

Figure~\ref{fit_ex} displays examples of spectra derived for the local areas of $30''\times30''$. Their spatial variation along the line connecting the Arches and Quintuplet clusters is shown in the upper panels, while that along the Galactic plane is in the lower panels. The shape of the spectrum varies significantly from area to area. In particular, the spectra of the regions around $(\ell,b)\simeq(0^{\circ}.13,0^{\circ}.0)$ exhibit a relatively strong hump around 80--90~cm$^{-1}$ (110--130~$\mu$m), as seen in the spectrum of the Arches cluster (Fig.~\ref{sed_all}b). Since some local spectra show unrealistically steep continuum slopes at wavenumbers smaller than 64~cm$^{-1}$ ($>$ 156~$\mu$m), probably due to calibration errors around the cut-on wavenumber of the detector responsivity that varies among the array pixels (\citealt{Mur10}), we additionally mask the spectral region at $<65$~cm$^{-1}$. We fit all the local spectra by the two dust modified blackbody components as used in the above fitting, plus the [O{\small III}] and [N{\small II}] emission lines, the OH absorption line, and the excess component. All the line components are assumed to be narrow lines, while the excess component is assumed to have a Lorentzian profile with the center variable between 80 and 95~cm$^{-1}$ and the width between 10 and 20~cm$^{-1}$. All the spectra are fairly well reproduced by the model; examples of the fitting results can be seen in Fig.~\ref{fit_ex}.

As a result, a local spectrum is decomposed into the three continuum components as well as the three line components at every spatial bin, and hence their spatial distributions are obtained. Each component is integrated with respect to the wavenumber; the total brightnesses of the warm and cold dust components are obtained by integration over the observed wavenumber range of 60--140~cm$^{-1}$. The distributions of the warm and cold dust components are shown in Figs.~\ref{fig5}a and b, while the distribution of the excess component is shown in Fig.~\ref{dustf}a. The results for the [O{\small III}] and [N{\small II}] lines are used in Fig.~\ref{fig4}. The OH absorption line flux is divided by the continuum intensity at the central wavenumber to derive its equivalent width, the distribution of which is shown in Fig.~\ref{dustf}b. The OH absorption due to the fundamental rotational mode becomes optically thick quite easily (\citealt{Goi02}), and therefore it might be saturated in high-resolution spectra of relatively strong regions, which would then give us only qualitative information. 

We find that the warmer dust emission is spatially correlated with the [O{\small III}] line emission (Fig.~\ref{fig5}a) and also with the [N{\small II}] line emission that is extended farther from the two clusters (Fig.~\ref{fig4}b). As can be seen in Fig.~\ref{dustf}, the distribution of the excess emission, which is apparently different from that of the warm dust emission, is notably similar to that of the OH absorption equivalent width. This spatial correlation, as well as the presence of their likely CO cloud counterpart as shown below, convinces us that the excess is not an artifact but a real spectral feature. In addition, the two independent detectors (i.e., SW and LW), both covering the excess component, consistently show a hint of the excess for the spectra indicative of its presence.

\begin{table*}
\caption{Results of the two-temperature modified blackbody model fitting to the spectra in Figs.~\ref{sed_all} and \ref{dustf2}.}
\label{table1}    
\centering         
\begin{tabular}{l c c c c} 
\hline\hline             
Spectrum & $T_{\rm w}$~(K) & $T_{\rm c}$~(K) & $\tau_{90\rm w}$\tablefootmark{a} & $\tau_{90\rm c}$\tablefootmark{a}\\   
\hline 
Figure~\ref{sed_all}a & 51~$\pm$~3 & 23.0~$\pm$~0.3 & (6~$\pm$~1)$\times10^{-3}$ & 0.33~$\pm$~0.01\\
Figure~\ref{sed_all}c & 51 (fixed) & 21.1~$\pm$~0.2 & (6.5~$\pm$~0.1)$\times10^{-3}$ & 0.54~$\pm$~0.02\\ 
Figure~\ref{sed_all}d & 51 (fixed) & 22.7~$\pm$~0.2 & (3.8~$\pm$~0.1)$\times10^{-3}$ & 0.34~$\pm$~0.01\\
Figure~\ref{dustf2} & 51 (fixed) & 19.7~$\pm$~0.2 & (5.2~$\pm$~0.2)$\times10^{-3}$ & 0.49~$\pm$~0.02, 0.10~$\pm$~0.02\tablefootmark{b}\\  
\hline  
\end{tabular}

\tablefoottext{a}{Optical depths at a wavelength of 90 $\mu$m for the warm and cold components. }
\tablefoottext{b}{The second cold component with the emissivity for the large graphite population and the same temperature as the first cold component (see text for details).}

\end{table*}

\section{Discussion}

\subsection{Warm dust component and its relation with molecular clouds}

\begin{figure*}
   \centering
   \includegraphics[width=0.45\textwidth]{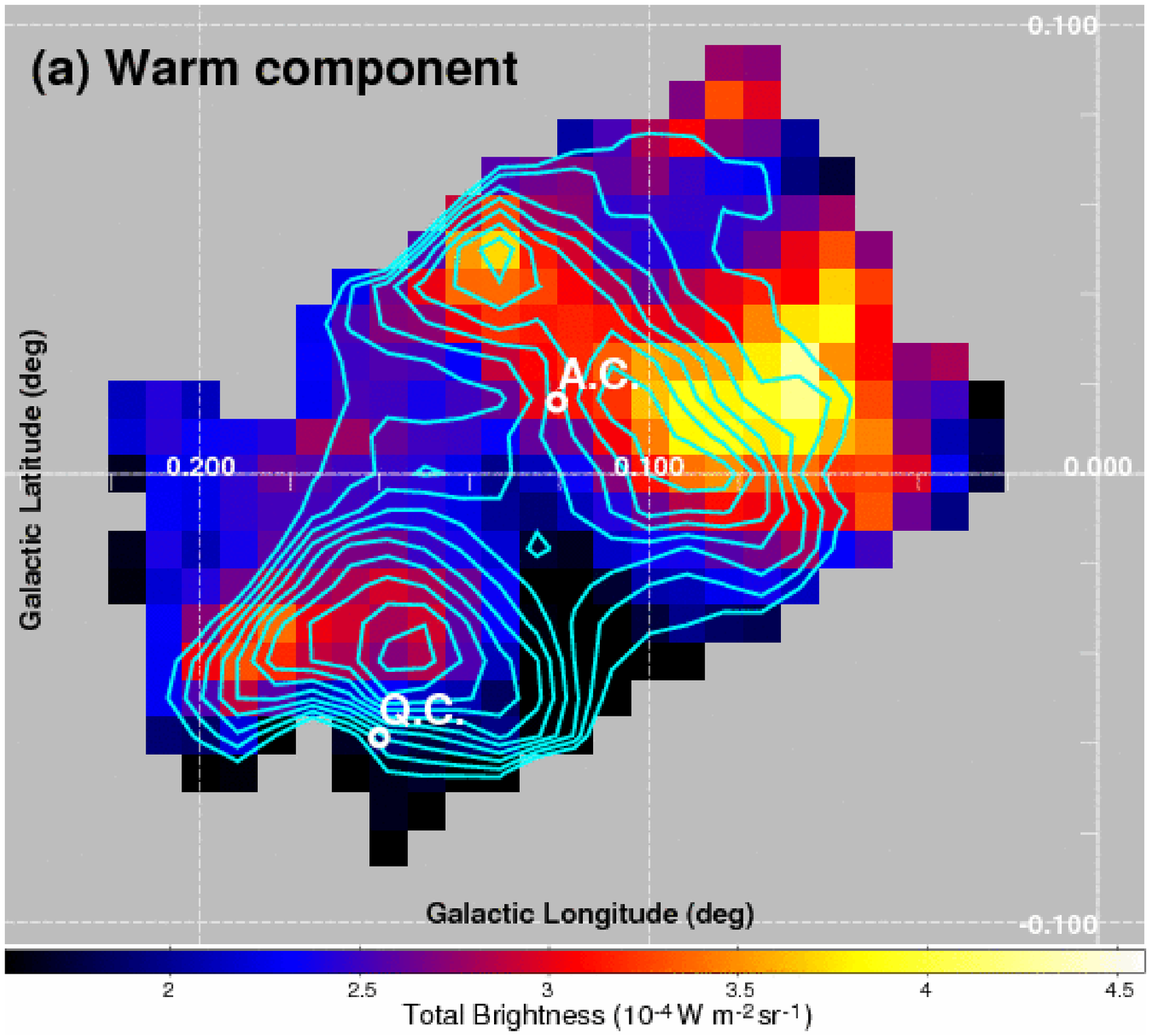}
   \includegraphics[width=0.45\textwidth]{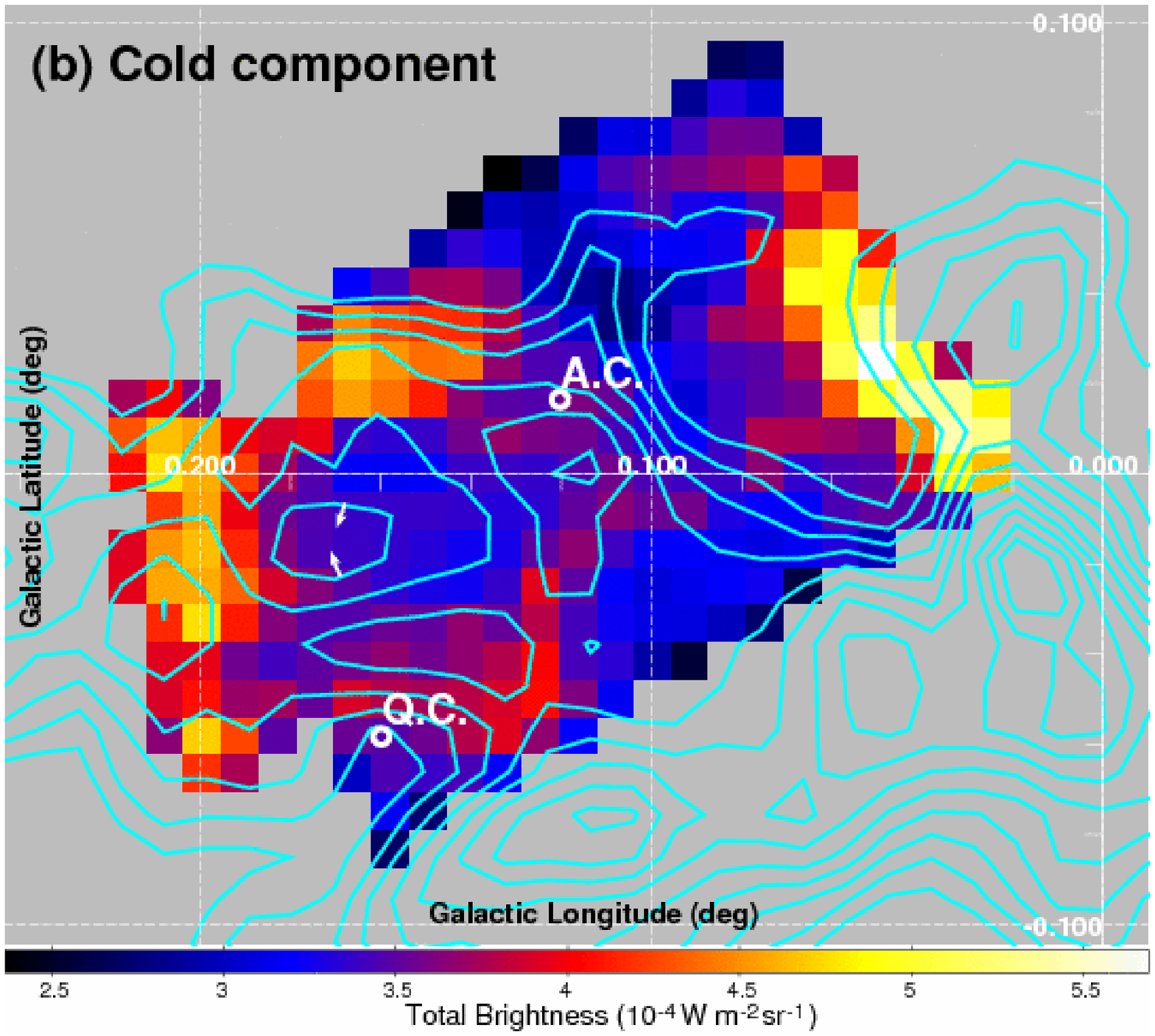}\\
   \includegraphics[width=0.45\textwidth]{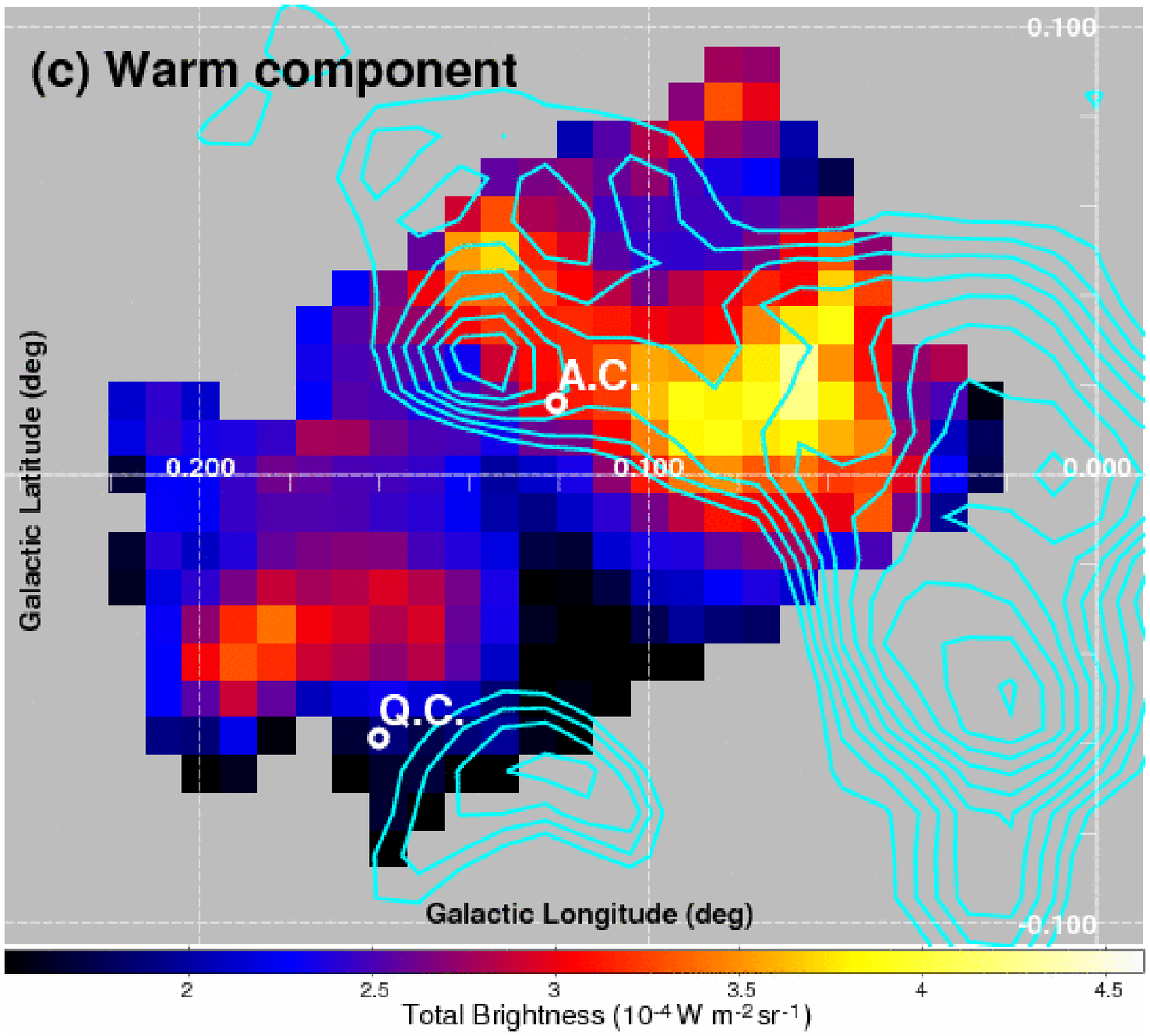}
   \includegraphics[width=0.45\textwidth]{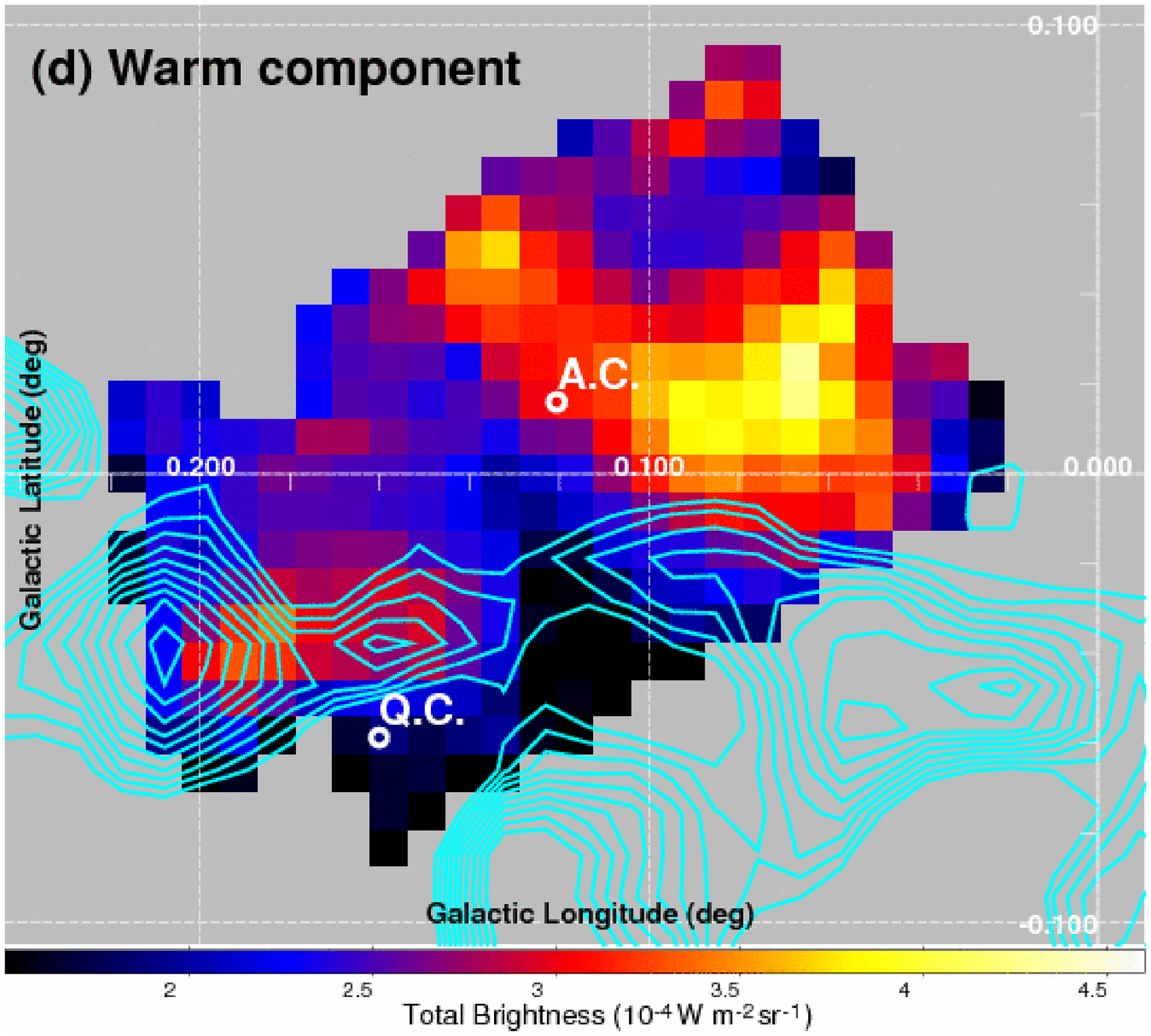}
      \caption{(a) Distribution of the total brightness of the warm dust component with the contour map of the [O{\small III}] line emission from the same data as in Fig.~\ref{fig4}b. The contour levels are linearly drawn from 1$\times$10$^{-6}$ to 3$\times$10$^{-6}$~W~m$^{-2}$~sr$^{-1}$. (b) Distribution of the total brightness of the cold dust component with the contour map of the $^{12}$CO J=1--0 emission integrated over the total velocity range ($V_{\rm LSR} = -220$ to $+220$~km~s$^{-1}$; Oka et al. 1998). The contours are linearly drawn from 690 to 1200 K km s$^{-1}$. (c) The contour maps of the $^{12}$CO J=1--0 emission integrated over the velocity range of $-30$ to $-10$~km~s$^{-1}$ and (d) that over $+20$ to $+40$~km~s$^{-1}$, both superposed on the same distribution map of the warm dust component as shown in panel (a). The contours are linearly drawn from 50 to 130~K~km~s$^{-1}$ for panel (c) and from 160 to 240~K~km~s$^{-1}$ for panel (d). }
         \label{fig5}
   \end{figure*}            

As shown in Fig.~\ref{fig5}a, the warm dust emission derived from the spectral decomposition is likely to be associated with the clouds influenced by the Arches and Quintuplet clusters. It is clear from the figure that the warm dust is more abundant near the Arches cluster than near the Quintuplet cluster. Using the optical depths $\tau_{\nu}$ derived from the spectral fitting, we estimated the mass of the warm dust existing around each cluster. For the square apertures of $1{'}.5\times1{'}.5$ used to create the spectra in Figs.~\ref{sed_all}c and d, $\tau_{\nu}$ at 90~$\mu$m is $6.5\times10^{-3}$ for the Arches cluster and $3.8\times10^{-3}$ for the Quintuplet cluster (table~\ref{table1}). Then, using the equation $M_{\rm dust}=\tau_{\nu}A/\kappa_{\nu}$, where $A$ is the physical area $\sim2\times10^{38}$~cm$^{2}$ with the distance to the Galactic center of 8.5~kpc and $\kappa_{\nu}$ is the mass absorption coefficient of 25~cm$^{2}$~g$^{-1}$ at 90~$\mu$m (\citealt{Dor95}), we obtain the warm dust masses of 26~M$_{\sun}$ and 15~M$_{\sun}$ for the Arches and the Quintuplet cluster, respectively. For comparison, we estimate the cold dust masses to be $2.1\times10^3$~M$_{\sun}$ and $1.3\times10^3$~M$_{\sun}$ with the same apertures for the Arches and the Quintuplet cluster, respectively. Since our far-IR coverage is not sensitive to the presence of even colder dust, they are rather lower limits to dust masses. Hence the warm dust makes only a negligible contribution to the total dust mass even toward the two active clusters, causing the clusters almost invisible in the far-IR dust continuum emission maps (Fig.~\ref{fig3}) and the brightness ratio map (Fig.~\ref{fig4}a).

The warm dust clouds are distributed anistropically around the Arches and the Quintuplet cluster, probably tracing the highly-ionized gas locally heated by intense stellar radiation from the two clusters. We then search for corresponding molecular clouds on the basis of the $^{12}$CO (J~=~1--0) map (Oka et al. 1998). As can be seen in Figs.~\ref{fig5}c and d, we find that CO clouds with the velocity ranges of $V_{\rm LSR} = -30$ to $-10$~km~s$^{-1}$ and $+20$ to $+40$~km~s$^{-1}$ show spatial correspondence to the warm dust distributions near the Arches and the Quintuplet cluster, respectively. It is therefore obvious that the warm dust clouds with the two clusters have different radial velocities. The velocity ranges are consistent with $\sim-30$~km~s$^{-1}$ and $\sim+40$~km~s$^{-1}$ (Serabyn \& G$\ddot {\rm u}$sten 1987, 1991; Lang et al. 2010) measured for the Arched Filaments and the Sickle H{\small II} region that are thought to be ionized by the Arches and the Quintuplet cluster, respectively. In other words, our results firmly support the connection between their connections. 

A larger amount of warm dust exists around the Arches cluster, while the [O{\small III}] emission is more prominent near the Quintuplet cluster. Thus there are differences in the properties of the ISM around the two clusters, presumably reflecting that the clusters differ in age; the Arches cluster is relatively young (2.0--2.5~Myr; Figer et al. 2002; Najarro et al. 2004) retaining a larger amount of the surrounding ISM, while the Quintuplet cluster is more evolved ($\sim$4~Myr; Figer et al. 1999) containing a variety of massive stars like luminous blue variables, one of which is the well-known Pistol star. 

In contrast, the cold dust component does not show any spatial correspondence with either the Arches or the Quintuplet cluster. In Fig.~\ref{fig5}b, we compare the distribution of the cold dust with that of the CO emission integrated over the total velocity range ($V_{\rm LSR} = -220$ to $+220$~km~s$^{-1}$). The figure shows that they have spatial distributions similar to each other. This is reasonable because the cold dust emission represents a significant fraction of the total gas amount integrated along the line of sight. We find that most of the cold dust is not likely to be associated with the two clusters. 

\subsection{Excess component over dust modified blackbody emission}

We find that there exists a significant excess on top of a modified blackbody continuum around 80--90~cm$^{-1}$ (110--130~$\mu$m), which probably represents a dust feature. 
From the spatial information, it is clear that this spectral component is not associated with either of the two clusters. By comparison with the CO map, a molecular cloud with velocities of $+70$ to $+90$~km~s$^{-1}$ is found to exhibit a good spatial correspondence to the excess component. Judging from this velocity range with the spatial position, the cloud seems to be part of the 100-pc ring revealed by Herschel around the Galactic center (\citealt{Mol11}), and in the foreground of the Arches cluster that is possibly located on the back side of the ring ($-30$ to $-10$~km~s$^{-1}$). From the negative to the positive longitude side of the Galactic center, the $+20$~km~s$^{-1}$ cloud, the $+50$~km~s$^{-1}$ cloud, and the above-mentioned $+70$ -- $+90$~km~s$^{-1}$ cloud are found next to each other, possibly forming part of the front side of the 100-pc ring. The position of Sgr A$^{*}$ is probably shifted from the center of the ring toward the front side (\citealt{Mol11}), which can be so close that the cloud may be significantly influenced by Sgr A$^{*}$.

With ISO, Kemper et al. (2002) detected a dust feature at 93~$\mu$m from the dust shells of evolved stars, which was identified as calcite (CaCO$_3$). Although the excess emission we detect appears in a similar far-IR region, it is a little shifted toward longer wavelengths ($110-130$~$\mu$m) and also certainly of interstellar origin. Hence, among likely interstellar grains, we suggest that the excess component may represent a graphite dust feature. Onaka \& Okada (2003) reported the detection of a similar broad feature at wavelengths from 80 to 140~$\mu$m in the ISO/LWS spectra of the diffuse emission from the Carinae Nabula and Sharpless 171. They pointed out a possibility that the feature originates from carbon onion grains. 
Graphite has an interband transition around 80~$\mu$m in the direction parallel to the graphitic plane (\citealt{Phi77}). However the interband transition is hardly visible for the ensemble of randomly oriented graphite spheres, because the absorption efficiency for the electric field parallel to the graphitic plane is much lower than that for the electric field vertical to the plane (\citealt{Dra84}). In the carbon onions, graphitic planes are curved, forming closed shells, and the optical properties in the direction parallel to the graphic plane interact with those in the perpendicular direction, which makes the interband feature become visible in the emissivity (\citealt{Hen93}). The position and profile of the interband feature are expected to depend significantly on assumed optical constants, especially the behavior of free electrons that depends on the temperature and the localization of $\pi$ electrons (\citealt{Tom01}). Therefore the peak position of the feature can be shifted to wavelengths longer than 100~$\mu$m (\citealt{Ona03}). 
Because it is suggested that harsh conditions in interstellar processes are favorable for the formation of onions (\citealt{Uga95}), our detection of a feature from the Galactic center, similar to that in the active star forming regions by Onaka \& Okada (2003), seems to be plausible.

As an alternative, we here propose a possibility of large graphite grains dominating in the far-IR emissivity. In the case of graphite grains larger than 0.5~$\mu$m, the magnetic dipole absorption, that is energy losses due to the eddy currents in the graphitic plane, becomes important, which can have wavenumber dependence reproducible of a spectral feature similar to the observed one (\citealt{Dra84}). Using the absorption efficiency $Q_{\rm abs}(a, \lambda)$ tabulated by B. T. Draine (\citealt{Lao93}) for the dust temperature of 25~K, where $a$ is the grain size, we calculated the total emissivity for two kinds of the size distributions for graphite dust: $n(a)da \propto a^{-3.5}da$ from $a_{\rm min}=0.001$~$\mu$m to $a_{\rm max}=0.25$~$\mu$m, and $n(a)da \propto a^{-2.5}da$ from $a_{\rm min}=0.001$~$\mu$m to $a_{\rm max}=4$~$\mu$m. The former is a standard distribution (\citealt{Mat77}), while the latter is the distribution strongly biased toward larger sizes. Then we fit the spectrum in Fig.~\ref{sed_all}c by a dust model consisting of two cold and one warm dust components, where the second cold component has the emissivity for the large graphite population, while the others have the emissivity for a standard grain size population. For simplicity, we assume that the two cold components have the same temperature but the fraction of the second to the first cold component and their common temperature is set to be free, while the temperature of the warm dust component is fixed at 51~K as above. As a result, the dust model including the large graphite grains reproduces the observed spectrum fairly well (Fig.~\ref{dustf2}), where the best-fit temperature of the cold dust components is $\sim$20~K (table~\ref{table1}). In terms of the dust mass, the additional large grain component contributes to about 5~\% of the total. 
Recently, Pagani et al. (2010) found the ubiquity of micron-sized dust grains in the dense ISM, using a so-called core-shine effect. Therefore it seems that grains much larger than standard dust grains can be present, but they must be mostly composed of graphite.

\begin{figure}
   \centering
   \includegraphics[width=0.45\textwidth]{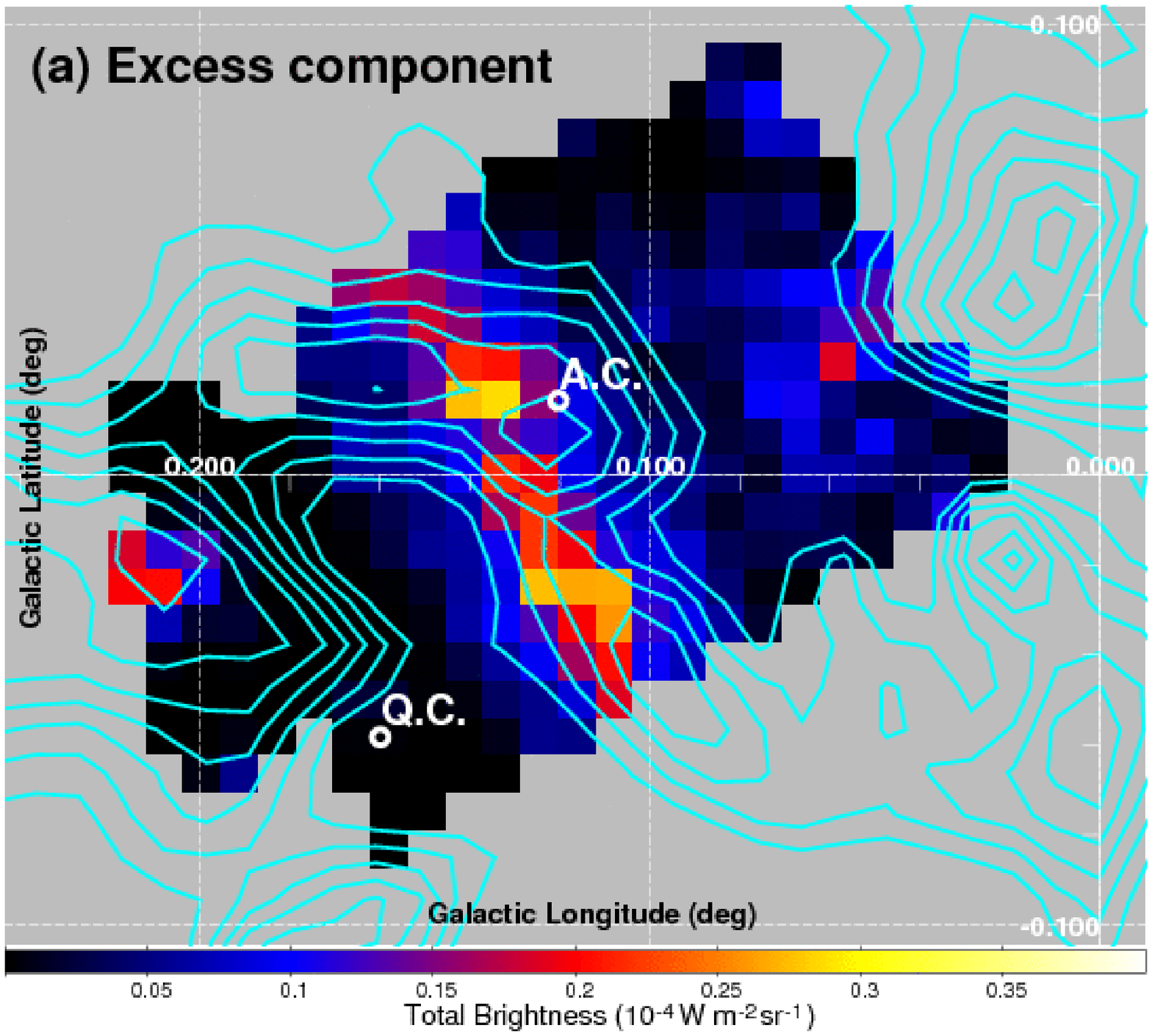}
   \includegraphics[width=0.45\textwidth]{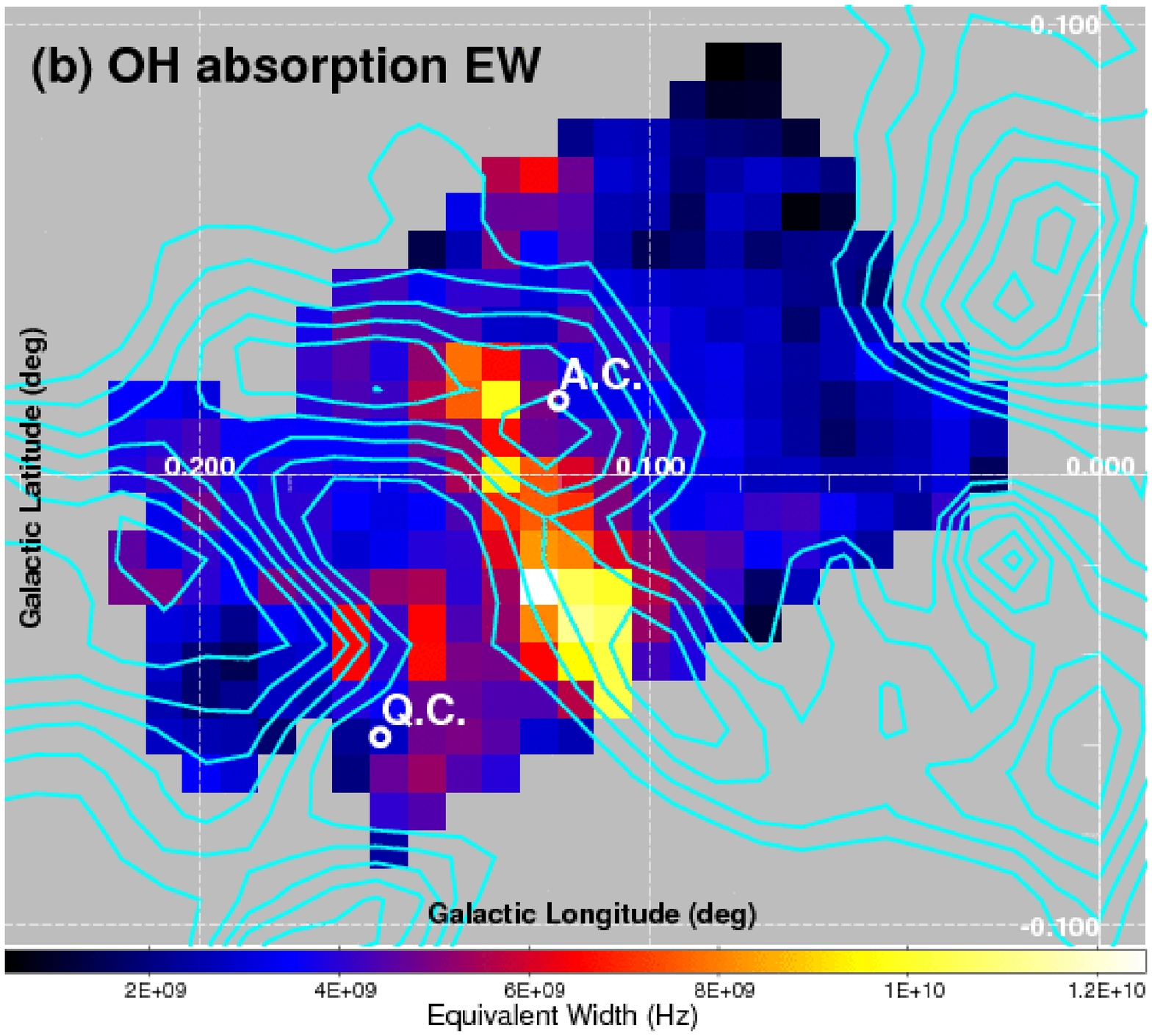}
      \caption{(a) Distribution of the total brightness of the excess component with the contour map of the $^{12}$CO(J=1--0) integrated over the velocity range of $+$70 to $+$90~km~s$^{-1}$. The contours are linearly drawn from 100 to 280~K~km~s$^{-1}$. (b) Distribution of the equivalent width of the OH 119~$\mu$m absorption line, which is derived by dividing the line flux by the continuum intensity at the central wavenumber. The contour map is the same as that in panel (a).  }
         \label{dustf}
\end{figure}

\begin{figure}
   \centering
   \includegraphics[width=0.45\textwidth]{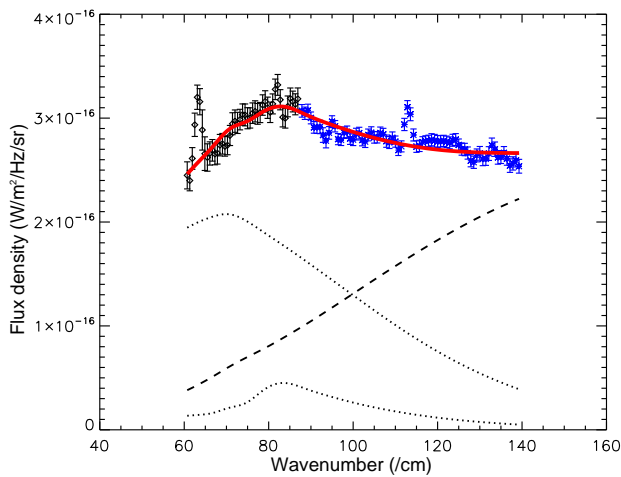}
   \caption{Same as Fig.~\ref{sed_all}c, but the spectrum is fitted by the warm and cold dust model including the second cold component with the large graphite population (see text for details). The two cold components and the warm component are indicated by the dotted lines and the dashed line, respectively.}
         \label{dustf2}
\end{figure}

It is also intriguing that the OH absorption is particularly enhanced in the region bright in the excess emission, as seen in Fig.~\ref{dustf}. The OH absorption and the excess emission are thus likely to originate in the same $+70$ -- $+90$~km~s$^{-1}$ cloud. The enhanced OH abundances are predicted in molecular shock regions (\citealt{Dra83}) or in the outer layers of photo-dissociation regions (PDRs, \citealt{Ste95}; \citealt{Goi04}).
Considering a possibility of dust shattering, the molecular shock would be hostile to the survival of the large grains, and therefore the PDR picture is more favorable in this case. Both scenarios for the dust feature suggest the importance of grains of graphite, although graphite materials are no longer believed to be major components of interstellar carbonaceous grains, but amorphous carbon materials (e.g., \citealt{Com11}). The detection of the excess feature from this region might be related with the (past) activity of Sgr A$^{*}$, because intense UV heating is needed to graphitize carbonaceous grains.   
  
\section{Summary}
We have presented the results of the far-IR FTS (60--140~cm$^{-1}$; $R=1.2$~cm$^{-1}$) spectral mapping of the Galactic center region of an area of about $10'\times10'$ to investigate the properties of interstellar dust around the massive star-forming clusters, Arches and Quintuplet. The maps of the continuum emission reveal similar spatial distributions at a range of wavenumbers (69--131~cm$^{-1}$), which do not clearly show any spatial correspondence to either the Arches or the Quintuplet cluster. There is no appreciable increase in the color temperature of the dust emission near the Arches and Quintuplet clusters, whereas strong [O{\small III}] line emission is clearly associated with the two clusters. Two dust modified blackbody components with different temperatures ($\sim$~50~K and $\sim$~20~K) can well reproduce the continuum shapes of most of the local spectra. Some spectra, however, reveal a significant excess on top of a modified blackbody continuum around 80--90~cm$^{-1}$ (110--130~$\mu$m), probably representing a dust feature. We find that the spatial distribution of the warmer dust component is well correlated with that of the [O{\small III}] line emission and hence the warm dust is likely to be associated with the highly-ionized gas locally heated by intense stellar radiation from the two clusters. The excess component exhibits an apparently different distribution from that of the warm dust component. 

We find that the CO clouds with different velocity ranges of $-30$ to $-10$, $+20$ to $+40$, and $+70$ to $+90$~km~s$^{-1}$ are spatially correlated with the warm dust component around the Arches cluster, that around the Quintuplet cluster, and the excess component, respectively. The first two velocity ranges agree with those measured for the Arched Filaments and the Sickle that are thought to be associated with the Arches and the Quintuplet cluster, respectively. Therefore we confirm their connections and successfully resolve the part of the CO clouds heated by the Arches and the Quintuplet cluster. The last velocity range suggests that the cloud associated with the excess component resides in the foreground of the Arches cluster. Together with the $+20$/$+50$~km~s$^{-1}$ clouds, it appears to form a front part of the 100-pc ring in the vicinity of Sgr A$^{*}$, where some grains might be highly processed to emit the far-IR feature. We discuss possible origins of the dust feature based on graphite materials: carbon onions and micron-sized graphite grains. We find that, at least, the dust model including the large graphite grains can reproduce the observed spectrum with the dust feature fairly well.

\begin{acknowledgements}
We thank all the members of the AKARI project. AKARI is a JAXA project with the participation of ESA. We use the archival $^{12}$CO (J~=~1--0) survey map of the Galactic center provided by T. Oka. et al. (1998). This research is supported by a Grant-in-Aid for Scientific Research No. 22340043 from the Japan Society for the Promotion of Science, and the Nagoya University Global COE Program, ``Quest for Fundamental Principles in the Universe: from Particles to the Solar System and the Cosmos'', from the Ministry of Education, Culture, Sports, Science and Technology of Japan. 
\end{acknowledgements}

\end{document}